\documentclass[11pt]{article}
\usepackage[a4paper,margin=0.96in]{geometry}
\usepackage[english]{babel}
\usepackage{algorithm}
\usepackage{algpseudocode}
\usepackage{amsmath}
\usepackage{cases} 

\usepackage{graphicx}
\usepackage[colorlinks=true, allcolors=blue, hypertexnames=false]{hyperref}

\usepackage[utf8]{inputenc} 
\usepackage{url}            
\usepackage{booktabs}       
\usepackage{amsfonts}       
\usepackage{nicefrac}       
\usepackage{microtype}      
\usepackage{xcolor}         

\usepackage{amsthm}
\usepackage{amsmath,amssymb}
\usepackage{comment}        
\usepackage[inline]{enumitem}  
\usepackage{bbm}
\usepackage{thmtools, thm-restate}

\usepackage{hyperref}
\usepackage{soul}

\newtheorem{theorem}{Theorem}[section]
\newtheorem{proposition}[theorem]{Proposition}

\newtheorem{lemma}[theorem]{Lemma}

\newtheorem{remark}[theorem]{Remark}
\newtheorem{definition}[theorem]{Definition}

\newtheorem*{definition*}{Definition}

\newtheorem*{theorem*}{Theorem}

\newtheorem*{lemma*}{Lemma}

\usepackage{bm}
\newcommand{\OPT}{\mathrm{OPT}}

\newcommand{\bz}{{\mathbf z}}

\newcommand{\bdelta}{\boldsymbol \delta}

\newcommand{\DACTE}{{\normalfont \text{DACTE}} }

\newcommand{\Lcal}{\mathcal{L}}

\newcommand{\Ocal}{\mathcal{O}}

\newcommand{\Tcal}{\mathcal{T}}

\newcommand{\Vcal}{\mathcal{V}}

\newcommand{\Xcal}{\mathcal{X}}

\newcommand{\Nbb}{\mathbb{N}}

\newcommand{\Rbb}{\mathbb{R}}

\newcommand{\bx}{{\boldsymbol x}}

\newcommand{\one}{\mathbbm{1}}

\newcommand{\final}[1]{{\color{red}}}

\newcommand{\ceil}[1]{{\left\lceil #1 \right\rceil}}

\newcommand{\Cost}{{\normalfont \texttt{Cost}}}

\newcommand{\OT}{\text{OT}}

\newcommand{\card}[1]{{\left| #1 \right|}}

\DeclareMathOperator*{\argmin}{arg\,min}

\usepackage{tikz}
\usetikzlibrary{trees,positioning}

\title{Asynchronous Collective Tree Exploration: \\
Communication Meets Randomness}
\author{Romain Cosson\thanks{Courant Institute, New York University, USA}
  \and
  Laurent Massoulié\thanks{Inria, Paris, France}}
\hypersetup{pdfauthor={Romain Cosson, Laurent Massoulié}}
\date{}

\begin{document}
\maketitle

\begin{abstract}
We study collective tree exploration, where a team of $k$ mobile agents must visit all nodes of an initially unknown tree using as few moves as possible. All agents start at the root and discover adjacent edges as they traverse the tree. Communication is \textit{distributed} in the sense that agents exchange information only by reading from and writing to whiteboards located at the nodes. Movements are \textit{asynchronous} in the sense that an adversary controls the agents' speeds at all times. Existing competitive guarantees for collective tree exploration are either distributed but synchronous, or asynchronous but centralized.
In contrast, we present a distributed asynchronous algorithm that explores any tree with $n$ nodes and depth $D$ in at most $2n+\Ocal(k 2^k D)$ moves, i.e., with regret linear in $D$. We also give a variant algorithm with a guarantee in $\Ocal(k/\log k)(n+kD)$, i.e., an $\Ocal(k/\log k)$ competitive ratio. 
We then show a lower bound on the competitive ratio of asynchronous collective tree exploration in $\Omega(\log^2 k)$, improving over the previous $\Omega(\log k)$ lower bound. Together, our algorithm and lower bound highlight an analogy between the role of communication in collective exploration and the role of randomness in online algorithms.  
\end{abstract}
\thispagestyle{empty}

\setcounter{tocdepth}{2}
\newpage
\tableofcontents
\thispagestyle{empty}
\clearpage
\pagenumbering{arabic}

\section{Introduction}
Collective tree exploration (CTE) is a well-studied problem modeling the exploration of an unknown environment by a team of $k$ mobile agents \cite{fraigniaud2006collective}. The problem formalizes the following question: \textit{Is depth-first search parallelizable?} Since its inception, two communication models have received most of the attention: the \textit{centralized} model---in which the agents can communicate at all times and are thus controlled by one central algorithm---and the \textit{distributed}\footnote{The agents are distinguishable so there is no need for symmetry breaking.} model---in which the agents may communicate only by reading and writing information on their current node.

Recent progress on the centralized model emerged from the study of a more general setting called asynchronous collective tree exploration (ACTE) \cite{cosson2024collective}. 
In ACTE, a team of $k$ agents initially located at the root of an unknown tree must visit all nodes in as few moves as possible. At each move, an adversary selects one agent that is allowed to: {\color{blue}(a)}~observe incident edges, {\color{blue}(b)}  communicate any information to other agents in the team, and {\color{blue}(c)}~move to one adjacent node.

An ACTE algorithm is said to have a regret of $f(k,D)$ if it explores any tree with $n$ nodes and depth $D$ in at most \[2n+kf(k,D)\quad \text{moves}.\] 
Translated back to synchronous collective tree exploration, where each of the $k$ agents gets one move at every time-step, such an algorithm explores any tree with $n$ nodes and depth $D$ in $\frac{2n}{k}+f(k,D)$ synchronous time-steps. The best regret guarantee to date is in $\Ocal(kD)$ \cite{cosson2024collective} and requires complete communication between the agents. 

An ACTE algorithm is also said to be $c(k)$-competitive if it explores any tree with $n$ nodes and depth $D$ in at most $c(k)(n+kD)$ moves. 

These measures of performance are standard and further motivated in Appendix \ref{ap:background-cte}.

\paragraph*{Main results.} In the first part of the paper, we initiate the study of the distributed model of asynchronous collective tree exploration (DACTE). In DACTE, the steps {\color{blue}(a)} and {\color{blue}(c)}  described above are unchanged, but step {\color{blue}(b)} is replaced by a local communication step. Specifically, the agents can communicate only by reading and writing information on fixed whiteboards at all nodes, like in \cite{fraigniaud2006collective}. To emphasize generality, we show DACTE naturally extends to agents moving continuously with adversarial speeds.   
We explain why the algorithm of \cite{fraigniaud2006collective} and all existing algorithms that handle distributed communication fail to achieve a non-trivial competitive ratio or a linear regret for DACTE. 
In contrast, we present an algorithm with a regret in $\Ocal(2^kD)$, and a variant algorithm with a competitive ratio in $\Ocal(k/\log k)$. We emphasize that our algorithm leads to the first linear-in-$D$ regret, even for the synchronous model of \cite{fraigniaud2004collective}.

In the second part of the paper, we prove a new lower bound $c(k)  =\Omega(\log^2 k)$ for the competitive ratio of ACTE. This immediately implies a regret lower bound in $\Omega((\log k)^2 D)$. The lower bound also applies to DACTE in which communications are restricted. 

Our regret guarantees are summarized in Table \ref{table:regret}. 
\begin{table}[H]
    \centering
    \begin{tabular}{||c | c c | c c||} 
     \hline
    Regret & Lower bound & Upper bound\\
    \hline
    Centralized algorithms (ACTE) & $\Omega(\log^2 (k) D)$ &$\Ocal(kD)$ \\
    Distributed algorithms (DACTE) & $\Omega(\log^2 (k) D)$ &$\Ocal(2^k D)$\\
    \hline
    \end{tabular}
\caption{Upper and lower bounds on the regret of asynchronous collective tree exploration. The centralized algorithm upper bound is due to \cite{cosson2024collective}, the three other entries are from this paper.}
\label{table:regret}
\end{table}

\paragraph*{Conceptual takeaway.} Our algorithm and lower bound rely on new connections with a celebrated problem in online algorithms called layered graph traversal, where \textit{randomness} plays a central role \cite{burley1996traversing,bubeck2023randomized}. Specifically, our \textit{distributed} algorithm is obtained by reducing DACTE to \textit{deterministic} layered graph traversal, whereas our \textit{centralized} lower bound is obtained by reducing \textit{randomized} layered graph traversal to ACTE. Both reductions are non-trivial and unidirectional (we believe there are no reciprocal reductions in either case). Nonetheless, we argue that they suggest a broader principle: that 
\textit{communication} in collective exploration plays a role analogous to \textit{randomness} in online algorithms. Motivated by this perspective, we conjecture that there is an exponential gap between distributed and centralized collective exploration and further conjecture that randomness is not a key resource for collective exploration. This is further substantiated in the conclusion.

\paragraph*{Related works.} Collective tree exploration was introduced by Fraigniaud, Gasieniec, Kowalski, and Pelc in 2004 \cite{fraigniaud2006collective}. It has received broad attention since, both in its centralized model of communication \cite{dynia2006smart,dynia2007robots,brass2011multirobot,ortolf2014recursive,brass2014improved,cosson2024breaking,cosson2024collective} and in its distributed model of communication \cite{fraigniaud2006collective,dynia2006power,brass2011multirobot,higashikawa2014online,dereniowski2015fast,cosson2023efficient}. The two models were defined in the original paper of \cite{fraigniaud2006collective}. The centralized setting is motivated by mobile computing, and AI, where the underlying tree can represent a combinatorial state space (e.g., a search tree). 
The distributed model is motivated by applications in parallel computing, such as garbage collection, whose goal is to identify all referenced objects during the execution of a program (reachability in pointer structures). For more concrete examples, we refer to \cite{rao1987parallel,ferguson1988distributed,singhal2019comparative, o2023dynamic} and references therein.

A central goal in collective exploration has been to investigate the competitive ratio. Classically, in the synchronous model, all $k$ agents are allowed to move at each round, and the competitive ratio of an algorithm is defined as $c(k) = \sup_{T} \nicefrac{\Cost(T)}{\text{OPT}(T)},$ where the supremum is taken over all trees $T$; here $\Cost(T)$ is the total number of rounds the algorithm requires to explore tree $T$ and $\text{OPT}(T)$ is the optimal offline cost.\footnote{An offline algorithm knows the tree layout but still has to traverse all edges.} Straightforward arguments (see Appendix \ref{ap:background-cte}) imply that $\text{OPT}(T) = \Theta(\nicefrac{n}{k}+D)$, thus, a collective tree exploration algorithm is order $c(k)$-competitive if and only if it explores any tree $T$ with $n$ nodes and depth $D$ in at most 
$\Ocal\left(c(k)(n/k +D)\right)$
synchronous rounds. Note that, under this definition, any algorithm whose cost is superlinear in $n$ or $D$, or that requires bounded values of $n$ or $D$ (even if this bound is in terms of $k$), has an infinite competitive ratio \cite{fraigniaud2006collective}.\footnote{This classical definition comes from online algorithms. For example, in the $k$-server problem, an algorithm whose competitive ratio depends on the length of the problem instance (i.e., the request sequence) is not considered competitive \cite{koutsoupias2009k}.}
Initially, \cite{fraigniaud2006collective} proposed a natural $\Ocal(k/\log k)$-competitive distributed synchronous algorithm. A number of other interesting guarantees followed, such as with superlinear cost \cite{brass2011multirobot,ortolf2014recursive}, and with linear cost assuming $n^{1+\epsilon}D\leq k$ \cite{dereniowski2015fast}. 

 Asynchronous collective tree exploration (ACTE) is the (a priori harder) problem in which an adversary selects one agent that may move at each step, and the cost of the team is the total number of moves required before the agents collectively explore the tree. We say that an algorithm for ACTE is $c(k)$-competitive if it explores any tree $T$ with $n$ nodes and depth $D$ in at most $\Ocal(c(k)(n+kD))$ moves. Asynchronous algorithms can be used in synchronous rounds, up to an arbitrary tie-breaking rule, with a speedup by a factor $k$ (because $k$ agents move at each round instead of one, see Lemma~\ref{lem:asynch-synch} for details). Thus, an order $c(k)$-competitive asynchronous algorithm induces an order $c(k)$-competitive synchronous algorithm, as defined above.

Recently, guarantees of the form $2n + f(k,D)$ were obtained for ACTE \cite{cosson2023efficient,cosson2024breaking,cosson2024collective}, translating to $2n/k + f(k,D)/k$ in the synchronous setting. The quantity $f(k,D)/k$ is referred to as the ``penalty'' or ``competitive overhead'' or ``regret'' since the other term in $2n/k$ is not improvable. Currently, the best regret guarantee is $\Ocal(kD)$ \cite{cosson2024collective}. This algorithm, when used with teams of $k'=\sqrt{k}$ agents, also yields the best competitive ratio to date, in $\Ocal(\sqrt{k})$ (Section \ref{sec:extensions} explains this simple reduction). The techniques in \cite{cosson2024collective} are inspired by online metric problems \cite{bansal2010metrical,buchbinder2014competitive,bubeck2023randomized, bubeck2022shortest}. Unfortunately, the approach cannot be distributed because it relies on online convex optimization at the team level. Furthermore, it uses the so-called `tree-mining game' \cite{cosson2024breaking}, which requires all-to-all communication in its definition.

Lower bounds on the competitive ratio of collective tree exploration have also been investigated. The first lower bound, due to \cite{fraigniaud2006collective}, is in $\Omega(1)$. It was then improved to $\Omega(\log k / \log \log k)$ by \cite{dynia2007robots,disser2017general}. This bound is easily pushed to $\Omega(\log k)$ against an asynchronous adversary. 
The absence of further progress on lower bounds echoes the long-standing randomized $k$-server conjecture, which claimed the existence of an $O(\log k)$-competitive algorithm for the randomized $k$-server problem, and which was recently refuted by \cite{bubeck2023randomized} which showed a lower bound of $\Omega(\log^2 k)$. In fact, we use their result to prove our own $\Omega(\log^2 k)$ lower bound.

Finally, we note that the broader problem of exploring an unknown environment with multiple agents is a recurring theme in computer science, with several potential emerging applications, 
such as robotics~\cite{brass2011multirobot,megow2012online,baligacs2023exploration}, parallel algorithms for branch-and-bound \cite{borst2025nearly}, multi-agent reinforcement learning \cite{blanke2023flex}, retrieval in knowledge graphs \cite{banerjee2022graph}, and parallelization of ``tree of thoughts'' paradigm for reasoning with large language models \cite{wei2022chain,yao2023tree,chen2023walking}. More broadly, we believe that online algorithms can help design simple rules for the interaction of autonomous intelligent agents \cite{jadbabaie2003coordination}.

\paragraph*{Outline of the paper.} In Section \ref{sec:lgt}, we provide essential background on the problem of layered graph traversal and then provide an overview of the techniques employed throughout the paper. In Section~\ref{sec:setting}, we introduce the setting of distributed asynchronous collective tree exploration (DACTE).   
In Section~\ref{sec:algo}, we introduce our exploration algorithm. In Section \ref{sec:lower-bound}, we prove a new lower bound on the competitive ratio of asynchronous collective tree exploration.

\section{Background and technical overview}\label{sec:lgt}

\subsection{Notation and definitions}
In this paper, somewhat unconventionally, we define trees as subsets of nodes, where nodes are defined as sequences of port numbers. The main benefit of this presentation is that it provides a unified framework for collective tree exploration and layered graph traversal, which will be exploited throughout the paper. 

\textit{Nodes.}  Let
$\Vcal = \cup_{d\in \Nbb}\Nbb^d$ be the set of all \textit{nodes},
with the convention that $\Nbb^0 = \{r\}$ represents the root. For a node $u=(u_1,\dots,u_d)\in\Vcal\setminus\{r\}$, $p(u) = (u_1,\dots,u_{d-1})\in \Vcal$ is the \textit{parent} of $u$ and $d(u)=d$ is the \textit{depth} of $u$. The list $(u_1,\dots, u_d)$ can be thought as a list of \textit{port numbers} leading from the root to node $u$. \textit{Ancestors} of $u$ are all the prefixes of $u$, including $u$ itself.
We write $v\succeq
u$ if $v$ is an ancestor of $u$ (or $u$ is a descendant of $v$). We denote by $A(u,v) \in \Vcal$ the \textit{lowest common ancestor} of $u$ and $v$, which equals the longest common prefix of $u$ and $v$. The distance between two nodes $u,v\in \Vcal$ is given by $d(u,v) = d(u)+d(v) - 2d(A(u,v))$. We observe that $(\Vcal,d(\cdot,\cdot))$ is a metric space.

\textit{Trees.} A (finite, rooted, ordered) \textit{tree} $T$ is a finite subset of $\Vcal$ satisfying:
$$u\in T\setminus\{r\} \implies p(u) \in T.$$
For any node $u\in T$, we denote by $C_u$ the list of children of $u$ in $T$, and by $c_u$ its cardinality, $c_u = \card{C_u}$. The edge set is $E = \{(p(u),u) : u\in T\setminus\{r\}\}$. 
The depth of the tree $T$ is $D = \max_{u\in T}d(u)$, and its size is $n = \card{T}$. 

\textit{Layers.} A layer $\Lcal\subset \Vcal$ is a subset of nodes that are incomparable for the ancestor partial order $\succeq$, i.e. $\forall u,v\in \Lcal: u\succeq v \implies u=v$ (a layer is also called an antichain in order theory). For two layers $\Lcal,\Lcal' \subset \Vcal$, we write $\Lcal\succeq \Lcal'$ if all nodes in $\Lcal'$ have an ancestor in $\Lcal$. We also define $\Tcal(\Lcal)$ the \textit{active subtree} associated with  layer $\Lcal$ which is composed of nodes that have a descendant in $\Lcal$, i.e., 
$$\Tcal(\Lcal) = \{u \in \Vcal : \Lcal_u\neq \emptyset\},$$
where
$\Lcal_u$ contains the descendants of $u$ that are in layer $\Lcal$.
When the layer is clear from context, we will denote $\Tcal(\Lcal)$ by $\Tcal$. A configuration on layer $\Lcal$ is denoted by $\bx \in \Xcal(\Lcal)$ and is a probability distribution on $\Lcal$. For a reason that will become clear in the definition of the transport cost below, it will be convenient to define a configuration on a given layer as an element of a larger set of all configurations $\Xcal$, which we now describe.

\textit{Configuration.} The set of all configurations, $\Xcal$, is defined by, 
$$\Xcal =  \left\{\bx \in \Rbb_+^\Vcal \text{ s.t. } x_r = 1 \text{ and } \forall u\in \Vcal: x_u \geq \sum_{\substack{v\in \Vcal \\  p(v)=u}} x_v \right\}.$$
It is easily verified that the transformation $: (x_u)_{u\in \Vcal} \rightarrow (x_u-\sum_{v: p(v)=u}x_v)_{u\in \Vcal}$, defines a one-to-one correspondence between configurations and finitely supported probability distributions over $\Vcal$. For some layer $\Lcal\subset \Vcal$, the set $\Xcal(\Lcal) \subset \Xcal$ is defined as the set of configurations $\bx$ with all their probability mass on $\Lcal$. Note that $\bx \in\Xcal(\Lcal)$ can be identified to a probability distribution over $\Lcal$, via the identity $\forall u\in \Vcal : x_u = \sum_{\ell\in \Lcal_u}x_\ell$ with $\sum_{\ell\in \Lcal} x_\ell = 1$. 

\textit{Transport Cost.}
For two configurations $\bx,\bx' \in \Xcal$, we define the transport cost between $\bx$ and $\bx'$ by
$$\OT(\bx,\bx') = \sum_{u\in \Vcal}|x_u-x'_u|.$$
It defines a distance on configurations, which coincides with the optimal transport distance between the underlying distributions in the metric space $(\Vcal,d)$ (see e.g., \cite{bubeck2021metrical}). We also define a related quantity, 
$\OT^{\uparrow}(\bx,\bx') = \sum_{u\in \Vcal}(x_u-x_u')^+$
which informally represents the transport cost from the distribution underlying $\bx$ to the distribution underlying $\bx'$ when only the movement from children to their parents is charged. We analogously define $\OT^\downarrow(\bx,\bx') = \OT^{\uparrow}(\bx',\bx)$. 

\textit{Other.} The logarithm $\log(\cdot)$ refers to the logarithm in base $2$, whereas $\ln(\cdot)$ refers to the natural logarithm. $\Nbb$ is the set of all integers. For any real $a\in \Rbb$, we write $a^+ = \max\{a,0\}$.

\subsection{Background on layered graph traversal}\label{sec:inputlgt}

\paragraph{Historical background.}
Layered graph traversal is a problem originating from the field of online algorithms which was introduced by Papadimitriou and Yannakakis in the early 1990s \cite{papadimitriou1991shortest}. The goal is for a single agent to traverse an unknown graph as fast as possible \cite{fiat1991competitive}, starting from an arbitrary source. 
The graph's nodes are organized into consecutive layers, which are revealed iteratively: when the agent reaches a given layer, it can see all nodes and edges up to that layer. 
The problem is typically parameterized by the width $w$, which corresponds to the maximum number of nodes in a layer. 
A particularly fascinating question in layered graph traversal is how \textit{randomness} affects the competitive ratio. Both the deterministic and randomized competitive ratio are now determined within a constant factor. The randomized competitive ratio scales in $\Theta(w^2)$ \cite{ramesh1995traversing,bubeck2022shortest,bubeck2023randomized}, and, following a recent improvement of \cite{coester2026chasing}, the deterministic competitive ratio scales in $\Theta(2^w)$ \cite{fiat1991competitive,burley1996traversing,coester2026chasing}. 
Layered graph traversal classically reduces to the special case of trees \cite{fiat1991competitive}, and, for convenience, in the rest of the paper, we refer to the problem as layered tree traversal or tree traversal. We now provide a formal definition of the problem.

\paragraph{Problem input.} For $w\in \Nbb$, a width-$w$ \textit{layered tree} is a monotone sequence $\Lcal(\cdot) = \Lcal(1)\succeq \Lcal(2)\succeq \dots$ of \textit{layers} (see Notation and definitions) satisfying $\forall t : \card{\Lcal(t)}\leq w$. The depth $D$ of the layered tree is the maximum depth of a node in $\Lcal(\cdot)$. 
Intuitively, a layered tree $\Lcal(\cdot)$ models the progressive unveiling of the \textit{underlying tree} $\cup_t \Tcal(\Lcal(t))$ over time. 

\paragraph{Deterministic algorithm.} A deterministic algorithm for tree traversal is a deterministic function $\ell(\cdot)$ mapping a sequence $\Lcal(\leq t)=(\Lcal(1),\dots,\Lcal(t))$ to an active leaf $\ell(t) \in \Lcal(t)$.\footnote{In a slight abuse of notation, we denote the $t$-th output of the algorithm by $\ell(t)$ instead of $\ell(\Lcal(\leq t))$.} The cost of algorithm $\ell(\cdot)$ on instance $\Lcal(\cdot)$ is equal to the sum of distances between its consecutive positions,
$$\Cost(\ell(\cdot),\Lcal(\cdot)) = \sum\nolimits_{t}d(\ell(t),\ell(t+1)).$$
A deterministic tree traversal algorithm $\ell(\cdot)$ is $c_w$-competitive if it satisfies that $\Cost(\ell(\cdot),\Lcal(\cdot)) \leq c_w D$ for all layered trees $\Lcal(\cdot)$ of width $w$ and depth $D$. Without loss of generality, we can restrict our attention to  lazy algorithms where $\ell(t)$ only moves if required; $\ell(t)\in \Lcal(t+1)\implies \ell(t+1)=\ell(t)$. 
\begin{theorem}[\cite{coester2026chasing}]\label{th:burley}
    There is a $\Ocal(2^w)$-competitive deterministic width-$w$ tree traversal algorithm. 
\end{theorem}

\paragraph{Fractional algorithm.} A fractional algorithm for tree traversal is a function $\bx(\cdot)$ mapping a partial input $\Lcal(\leq t)$ to a distribution over active leaves $\bx\in\Xcal(\Lcal(t))$. We shall denote its output at time $t$ by $\bx(t)$ instead of $\bx(\Lcal(\leq t))$ for simplicity. The cost of a fractional algorithm $\bx(\cdot)$ on the instance $\Lcal(\cdot)$ is equal to the optimal transport cost between consecutive distributions,
$$\Cost(\bx(\cdot),\Lcal(\cdot)) = \sum\nolimits_{t}\OT(\bx(t),\bx(t+1)).$$
A fractional algorithm $\bx(\cdot)$ is $c_w$-competitive if it satisfies
    $\Cost(\bx(\cdot),\Lcal(\cdot)) \leq c_w D$
for any tree traversal instance $\Lcal(\cdot)$ of width $w$ and depth $D$. Classically, fractional algorithms are equivalent to randomized algorithms because a fractional configuration is exactly the distributional description of a random trajectory (see e.g. \cite{bubeck2022shortest} for a formal justification). 
\begin{theorem}[\cite{bubeck2023randomized}]\label{lb:bubeck} The competitive ratio of width-$w$ fractional tree traversal satisfies 
$c_w=\Omega(w^2)$.
\end{theorem}

\subsection{Technical overview}\label{sec:overview}
\paragraph{First part of the paper: distributed exploration and algorithm.} The first part of the paper is devoted to defining distributed asynchronous collective tree exploration (DACTE) and presenting an algorithm for this new setting. The definition of DACTE follows naturally from previous works. We take some new simple conceptual steps towards increasing the generality of collective tree exploration, such as the extensions to weighted trees (Lemma~\ref{lem:weighted-edge}) and to continuous moves (Lemma \ref{lem:continuous}). We also explain why previous distributed algorithms fail to provide competitive guarantees for DACTE (Appendix \ref{sec:failure}). We then turn to our distributed algorithm. We note that, unlike previous works on asynchronous collective tree exploration, we cannot rely on the so-called ``tree-mining game'' \cite{cosson2024breaking}, 
because it uses complete communications between the agents. Therefore, we return to the elementary notion of ``locally-greedy'' algorithms, i.e., exploration algorithms where a moving agent always traverses an adjacent unexplored edge if possible and moves towards a ``target'' otherwise. Such algorithms are efficient whenever the total movement of the agents' targets is limited. We show that one can use a deterministic layered tree traversal algorithm $\ell(\cdot)$ to elect a sequence of agent targets $S = v^1,v^2,\dots$ by defining a tree traversal instance $\Lcal(\cdot)$ during the exploration. At some step $h\in \Nbb$ of the algorithm, $\Lcal(h)$ will represent (a proxy of) the set of the minimal elements for the partial order $\preceq$ of the discovered nodes that are adjacent to unexplored edges. The main difficulty is that, due to the nature of the distributed communications, no single agent has a complete representation of the entire subtree explored by the team. 
A concise (informal) description of the algorithm is given in Section~\ref{sec:informal}. The remainder of the effort is therefore spent on demonstrating that this algorithm can be formally implemented using distributed communications. The algorithm is formally stated in Appendix~\ref{sec:formal-desc}, where we carefully specify all variable updates from an agent's point of view. The detailed analysis of the algorithm is provided in Appendix \ref{sec:analysis}.

\paragraph{Second part of the paper: lower bound.} The second part of the paper is devoted to showing a new lower bound on asynchronous collective tree exploration. It utilizes the recent breakthrough of \cite{bubeck2023randomized} (best paper award at STOC 2023), which disproves the randomized $k$ server conjecture and provides a lower bound of $\Omega(w^2)$ for the competitive ratio of width-$w$ (fractional) tree traversal. Our proof exploits their lower bound via a new reduction from fractional tree traversal to ACTE.

It is not hard to show that a guarantee in $c(k)(n+kD)$ for ACTE induces a tree traversal algorithm $\bx(\cdot)$ satisfying
    \begin{equation}\label{eq:lb-sketch}
        \Cost(\bx(\cdot),\Lcal(\cdot)) \leq \frac{c(k)}{k}L + c(k) D,
    \end{equation}
where $L$ is the size (number of nodes) of the tree underlying $\Lcal(\cdot)$. This is seen by considering $k$ agents exploring the tree underlying the instance $\Lcal(\cdot)$, interpreting them as a probability distribution over $k$ points (with multiplicity allowed), with an adversary forcing the agents to move from one layer to the next sequentially (Lemma~\ref{lem:acte-ltt}). Yet, this does not yield a competitive algorithm for tree traversal due to the first term in \eqref{eq:lb-sketch} that depends on $L$. Note that, for now, Equation \eqref{eq:lb-sketch} uses no assumption on the width of $\Lcal(\cdot)$. The main idea of the proof (Proposition~\ref{prop:main} in Section~\ref{sec:lower-bound}) is that we can always remove the term in $L$ if the width $w$ of the tree traversal instance $\Lcal(\cdot)$ satisfies $\frac{c(k)}{k}\leq \frac{1}{w}2^{-w-1}$. Therefore, under this condition, we have a reduction of tree traversal to ACTE, and we can invoke the lower bound for the former \cite{bubeck2023randomized} to obtain a lower bound for the latter.

We now explain the idea of Proposition \ref{prop:main}, where we cancel a term in $\epsilon L$ in the right-hand side of Equation~\eqref{eq:lb-sketch} for layered trees of width $w$, where $\epsilon = \frac{1}{w}2^{-w-1}$. Our goal is to use $\bx(\cdot)$ to define another tree traversal algorithm $\bz(\cdot)$ that may have a higher cost in terms of $D$, but has a better guarantee in terms of $L$. A naive attempt would be to define $\bz(\cdot)$ by $z_\ell(t) = x_\ell(t)-\epsilon$, in the hope to cancel a cost in $\epsilon$ for every new node. But
this idea has several issues, two of which are major: (a) small changes in the shape of $\Tcal(t)$ (such as a leaf forks) would lead to large movements of $\bz(t)$, (b) the algorithm $\bz(\cdot)$ is not well-defined because its total mass varies with $\card{\Lcal(t)}$. Instead, we draw inspiration from the technique used in \cite{cosson2024barely} to study the role of randomness in metrical task systems. We define $\bz(\cdot)$ as follows
\begin{equation}\forall t: \bz(t) \in \argmin_{\bz\in \Xcal(t)} \OT^\uparrow(\bz(t-1),\bz)+D(t,\bx(t),\bz),\label{eq:first}
\end{equation}
where $D(t,\cdot,\cdot)$ is a time-dependent potential (which has a simple expression) that penalizes configurations satisfying $z_\ell \geq x_\ell(t)-\epsilon'$ for some well-chosen $\epsilon'$. The resulting algorithm $\bz(\cdot)$ has two desirable properties: the first is that it has a limited movement cost (thanks to the term in $\OT^\uparrow(\bz(t-1),\bz)$); the second is that $z_\ell(t) \leq (x_\ell(t)-\epsilon')$ (thanks to the potential). 
The proof differs largely from existing techniques. In particular, the potential is time-dependent via another fractional algorithm $\bdelta(t)$, which represents a probability distribution on $\Lcal(t)$ induced by a random depth-first search. 
We believe the approach may be of broader interest for online metric problems, especially when the underlying metric space varies or is infinite. 

\section{Results and extensions}\label{sec:setting}
\subsection{Model and results}\label{sec:model}
We fix an unknown tree $T$, and,
following~\cite{fraigniaud2006collective}, we assume that every node of $T$ contains a whiteboard of unbounded size that can be used to store information. Each of the $k\in \Nbb$ agents also has infinite internal memory, as well as the ability to read and write on the whiteboards and to perform any computation. Exploration is decomposed into discrete moves. Initially, at $t=0$, all agents/robots are located at the root of the unknown tree. At each $t\in \Nbb$, an omniscient adversary\footnote{The adversary knows the exploration algorithm, the tree, and the memory and position of all agents.} chooses one robot $i(t)\in [k]$ which may move. A move of a robot consists of the following consecutive instantaneous steps:
\begin{enumerate}[label=\alph*), itemsep=-0.1em]
    \item The robot observes the list of incident edges\footnote{Note that nodes are identified by a sequence of port numbers, see notations and definitions section.}; \label{step2}
    \item The robot may read from and write to the local whiteboard at its position;\label{step3}
    \item The robot moves to one adjacent node of its choice.\footnote{Note that the robots do not read or write at their destination.}\label{step4}
\end{enumerate}
Exploration terminates at the first $t\in \Nbb$ when all nodes have been visited by at least one robot. In Section~\ref{sec:algo}, we provide an algorithm with the following guarantee.
\begin{theorem}\label{th:redtm} There exists a $k$-DACTE algorithm exploring any tree with $n$ nodes and depth $D$ in at most $2n+\Ocal(k 2^k D)$ moves, and a variant of this algorithm that is $\Ocal(k/\log k)$-competitive.
\end{theorem}
Our lower bound, which is proved in Section~\ref{sec:lower-bound}, concerns the general setting of asynchronous collective tree exploration (ACTE), where step \ref{step3} is replaced with one round of all-to-all communication among the robots. It also applies to the distributed setting (DACTE). 
\begin{restatable}{theorem}{thlb}\label{th:lb}
    The competitive ratio of $k$-ACTE satisfies $c(k)=\Omega(\log^2 k)$.
\end{restatable}

\paragraph{Terminology.} The robots in the tree are indexed by $i\in [k] = \{1,\dots,k\}$. We say that a node $u\in T$ has been \textit{visited} when the adversary has activated a robot located at $u$. Therefore, the root is visited after the first robot is activated. For $c\in C_u$ a child of $u$ in $T$, we say that the edge $e=(u,c)$ is explored as soon as some robot located at $u$ has started to traverse that edge (possibly, this robot has not yet visited $c$). Otherwise, we say that $e$ is unexplored. Since edges are always discovered from their highest endpoint, we can assume, without loss of generality, that the information of whether an adjacent edge is explored is always available to a robot. 

\subsection{Simple extensions}\label{sec:extensions}
\paragraph*{Converting a regret into a competitive ratio.} The following lemma generalizes a standard argument. 
\begin{lemma}\label{lemma:regrettoratio}
    If there exists a \DACTE algorithm with a guarantee in $f(k,n,D)$, then, for any $k'\leq k$, there exists a \DACTE algorithm with a guarantee in $\ceil{k/k'}f(k',n,D)$.
\end{lemma}
\begin{proof}
 Consider the algorithm that divides the $k$ explorers into $\ceil{k/k'}$ teams of $k'$ robots (or fewer), without any interaction between the teams (robots ignore the messages left by other teams). By the generalized pigeonhole principle, after $\ceil{k/k'}\times M$ moves, at least one of the teams has received $M$ moves, so $M\leq f(k',n,D)$ at completion.
\end{proof}
This lemma turns our regret guarantee in $2n + \Ocal(k 2^k D)$ into the variant guarantee in $\Ocal(k/ \log k)(n +kD)$, i.e., $\Ocal(k/\log k)$-competitive. Indeed, consider the algorithm obtained by using Lemma \ref{lemma:regrettoratio} with $k' = \ceil{\ln k}$. Letting $f(k,n,D) = \Ocal(n+k 2^kD)$, we have that $f(k',n,D) = \Ocal(n+(\ln k)^2\exp(0.7\ln(k))D)=\Ocal(n+kD)$, yielding the desired guarantee.

\paragraph*{Weighted trees.} We now consider the problem where each edge $e$ has a given weight $w_e>0$ that is observed by any agent that is adjacent to that edge. The cost of traversing an edge $e$ is equal to $w_e$, and the team's goal is now to explore the entire tree while incurring a limited total cost. We denote by $L$ the sum of all edge weights in the tree and by $D$ the tree depth, defined as the maximum (weighted) distance of a node to the root. We assume that the edge weights are integral multiples of some known small real $a>0$. If edge weights are arbitrary reals in $\{0\}\cup [\epsilon,+\infty]$, they can simply be rounded to their closest value in $a\Nbb$, for $a<<\epsilon$. The following lemma is proved in Appendix \ref{ap:lem-model}. It is simply obtained by decomposing all edges of the tree in sub-segments of length $a$. 

\begin{restatable}{lemma}{lemweighted}
\label{lem:weighted-edge}
An asynchronous exploration algorithm that finishes in $f(k,n,D)$ moves in the unweighted setting entails an asynchronous exploration algorithm that explores with a cost of $a\times f(k,L/a,D/a)$ in the weighted setting, where edge weights are in $a\Nbb$. This quantity is, therefore, equal to $f(k,L,D)$ if $f(\cdot,\cdot,\cdot)$ is linear in its second and third arguments. 
\end{restatable}

\paragraph*{Continuous speeds.} The above presentation is discrete since the robots are required to move either by one complete edge or not to move at all.  Instead, we now assume that the adversary controls the (continuous) speeds of the robots, with robots possibly getting blocked inside an edge. The cost of the team is defined as the total distance traveled by the agents before the exploration terminates. 
The following lemma is proved in Appendix \ref{ap:lem-model}.
\begin{restatable}{lemma}{lemcontinuous}
\label{lem:continuous}
    An exploration algorithm with a guarantee in $f(k,n,D)$ in the discrete move model entails an exploration algorithm with a guarantee in $f(k,n,D)$ in the continuous move model. 
\end{restatable}
The above extensions motivate the description of the problem in the abstract, where we stated that the adversary controls the continuous speeds of all robots at all times. 
In the rest of the paper, we adopt the problem definition given in Section \ref{sec:model}.

\section{A distributed algorithm}\label{sec:algo}
We start with an informal (centralized) description of our algorithm in Section \ref{sec:informal}, and we defer the formal (distributed) algorithm description to Section \ref{sec:formal-desc}. The analysis (correctness and competitive guarantee) of our algorithm is given in Section \ref{sec:analysis}.

\subsection{Locally-greedy algorithms}\label{sec:loc-greedy}
We recall the basic definition of locally-greedy algorithms. 

\begin{definition}[Locally-Greedy Algorithm]\label{def:locally-g}
    A locally-greedy algorithm with targets is such that each robot $i$ maintains at all times a target node $v_i(t)$ and the moving robot $i(t)$ satisfies,
\begin{enumerate}[label=R\arabic*.,itemsep=-0.3em, topsep=-0.2em]
    \item \textbf{if} adjacent to an unexplored edge, \textbf{then} the robot traverses one such edge; \label{it: 1}
    \item \textbf{else} the robot moves towards its target $v_{i(t)}(t)$ ; \label{it: 2}
    \item The robot does not stay at its current position. \label{it: 3}
\end{enumerate}
\end{definition}
In short, an algorithm is locally-greedy if robots always prefer to traverse an unexplored edge when one is adjacent. An equivalent perspective on rule \eqref{it: 3} is that the value of the target $v_{i(t)}(t)$ must have been updated at the beginning of move $t$ if the following condition was raised: 
\begin{enumerate}[label=C.]
    \item The robot $i(t)$ is located at its target and is not adjacent to an unexplored edge. \label{it: change_anchor}
\end{enumerate}
Locally greedy algorithms are entirely defined by the update rule over targets. Furthermore, if the movement of the targets is limited, they enjoy good guarantees, as explained in the following proposition, whose proof is recalled in Appendix \ref{ap:lem-model}.   
\begin{restatable}[\cite{cosson2024breaking}]{proposition}{proptargets}\label{prop:targets} After $M$ moves of a locally-greedy algorithm with targets, the number of different edges that have been explored by the algorithm is at least,
 \begin{equation*} \frac{1}{2}\left(M - \sum_{i\in [k]}\sum_{t< M} d(v_i(t),v_i(t+1))\right).\end{equation*} 
\end{restatable}

\subsection{Informal description and analysis}\label{sec:informal}
The DACTE algorithm we propose is an instance of a locally-greedy algorithm with targets. Initially, all robots are targeted at the root. As explained in Section~\ref{sec:loc-greedy}, the target of a robot is updated whenever the condition \eqref{it: change_anchor} is met.

Over the course of the algorithm, all robots will use the same sequence of targets $S = v^1, v^2, \dots$ in the same order, although the robots will not have the same target at the same instant, with some robots being ahead of others. The robot that defines the most advanced target is temporarily referred to as the \textit{leader}. Once defined, the value of target $v^{h+1}$ and the path leading to it from $v^h$ will always be written on the node $v^h$. This allows robots with outdated targets (\textit{followers}) raising condition \eqref{it: change_anchor} at $v^h$ to update their target to the more recent target $v^{h+1}$. Note that the identity of the leader and followers may change over the course of exploration.

We now explain what happens when condition \eqref{it: change_anchor} is raised by the leader, for which the target $v^h$ has no successor. 
All the quantities used in this process will be registered on the whiteboard at $v^h$. 
For each robot $i\in [k]$, the leader defines $c(i,h+1)$ to be the child of $\{v^1,\dots,v^h\}$ under which the robot indexed by $i$ is currently exploring (if any) and sets $\Lcal(h+1) = \{c(i,h+1) ~s.t.~ i\in [k]\}$. In reality, $c(i,h+1)$ cannot be computed by the leader with distributed communications, and the real definition of $\Lcal(\cdot)$ will differ in the formal algorithm description in Appendix~\ref{sec:formal-desc} --- our purpose being to convey the spirit of the algorithm. 
We will show that at this point of exploration, all unexplored nodes must be descendants of $\Lcal(h+1)$ (see Claim 5 of Appendix \ref{sec:analysis} for more details). Thus, if $\Lcal(h+1)$ is empty, the exploration is completed. Otherwise, the leader uses a deterministic tree traversal algorithm $\ell(\cdot)$ on $\Lcal(\leq h+1) = (\Lcal(1), \dots,\Lcal(h+1))$ to define $v^{h+1} = \ell(h+1)$.
Finally, the leader updates its target to $v^{h+1}$, and performs one move in that direction. 

The high-level analysis in Appendix \ref{sec:analysis} is as follows: we first show that the distributed algorithm defined in Appendix \ref{sec:formal-desc} is locally-greedy (Claim 1) with all robots using the same sequence of targets $S = v^1,v^2,\dots$ (Claim 2). In light of Proposition \ref{prop:targets}, this proves that the number of moves before all $n-1$ edges are explored is at most
$2n + k\sum_{h} d(v^h,v^{h+1})$.
Then, we prove that despite distributed communications, the sequence $S = v^1,v^2,\dots$ comes from a valid layered graph instance $\Lcal(\cdot)$ (Claim~3 and Claim~4) of width $w=k$ (Claim~6) contained in a subtree of the underlying unknown tree, and that the instance does not finish before the tree is fully explored (Claim~5). In light of Theorem \ref{th:burley}, we have $\sum_h d(v^h,v^{h+1})\leq c_k D$, where $c_k = O(2^k)$. Therefore, the unknown tree is explored in at most 
\begin{equation*}2n + k\sum\nolimits_{h} d(v^h, v^{h+1}) = 2n+ O(k2^kD) \quad \text{moves.}\end{equation*}
An illustration of the algorithm is provided in Figure \ref{fig:illustration-algo}, in Appendix \ref{sec:formal-desc}.

\section{An improved lower bound}\label{sec:lower-bound}
\subsection{ACTE and tree traversal}
We start by introducing a new type of guarantee for tree traversal.
\begin{definition}
Let $a,b\in \Rbb^+$ be two reals. A fractional tree traversal algorithm $\bx(\cdot)$ is said to have $(a,b)$ overhead if it satisfies on any tree traversal instance $\Lcal(\cdot)$,  
\begin{equation}\label{eq:assumption}
        \Cost(\bx(\cdot),\Lcal(\cdot)) \leq aL +bD ;
\end{equation}
where $D$ and $L$ are the depth and length (number of nodes) of the tree underlying $\Lcal(\cdot)$. 
\end{definition}
The following result is a simple observation (see Appendix \ref{ap:lem-lb}).
\begin{restatable}[\cite{cosson2024collective}]{lemma}{lemacte}\label{lem:acte-ltt} Let $k\in \Nbb$. For any $k$-ACTE algorithm requiring at most $c(k)(n+kD)$ moves on trees with size $n$ and depth $D$, there is a fractional tree traversal algorithm $\bx(\cdot)$ that has $(c(k)/k, c(k))$ overhead. 
\end{restatable}
Our lower bound then follows from the following statement, which is the main technical result of this part of the paper, and which we prove in Section \ref{sec:mainprooflb}.
\begin{proposition}\label{prop:main}
    Let $w\in \mathbb{N}$ let $a<1/(w2^{w+1})$ and $b\in \Rbb^+$. If there is a fractional tree traversal algorithm 
    with $(a,b)$ overhead, 
    then, there exists a $(16b{}+3)$-competitive width-$w$ fractional tree traversal algorithm.
\end{proposition}
Before the proof of this proposition, we show how it implies Theorem \ref{th:lb}.
\thlb*
\begin{proof}[Proof of Theorem \ref{th:lb}.] We now explain why Proposition \ref{prop:main} entails our lower bound $c(k)=\Omega(\log^2 k)$ where $c(k)$ refers to the competitive ratio of $k$-ACTE. The proof uses a reduction from tree traversal to ACTE.

We first prove the lower bound on integers of the form $k\in \{w^3 2^{w+5} : w\in \Nbb\}$, and then extend it to all $k\in \Nbb$ by a simple monotonicity argument.\footnote{Specifically, for $k\in \Nbb$, denote by $k'$ the largest integer in $\{w^3 2^{w+5} : w\in \Nbb\}$ that is smaller than $k$. Noticing that $k'\geq k/16$, we have $c(k)\geq c(k')/16$. This is because an ACTE algorithm for $k\geq k'$ robots can be used to define an ACTE algorithm for $k'$ robots by simulating an adversary that leaves $k-k'$ robots idle.} Assume that $k$ is of the form $k = w^3 2^{w+5}$, for some integer $w\in \Nbb$. Applying Lemma~\ref{lem:acte-ltt}, there exists a tree traversal algorithm with overhead $(c(k)/k,c(k))$. Then, either (1) $c(k)/k \mathrel{<} 1/(w2^{w+1})$, which yields a $(16 c(k){}+3)$-competitive algorithm for width-$w$ tree traversal, implying that $16c(k){}+3\geq cst\cdot w^2$ for $cst$ the universal constant in the lower bound of tree traversal (Theorem~\ref{lb:bubeck}, \cite{bubeck2023randomized}) --- or (2) $c(k)/k \geq  1/(w2^{w+1})$, which rewrites as $c(k) \geq 8 w^2$. In both cases, we have shown that $c(k)\geq cst\cdot w^2$ for a universal constant $cst$, and we can conclude by noting that $w\geq (\log_3 k - 5)/2$. 
\end{proof}

\subsection{Proof of Proposition \ref{prop:main}}\label{sec:mainprooflb}
\begin{proof}[Proof of Proposition \ref{prop:main}]We now prove the main technical result. We refer to Section \ref{sec:overview} for an overview of the proof.
\paragraph{Preliminaries.} In this proof, $w\in \Nbb$ is fixed. We shall, without loss of generality, restrict our attention to instances $\Lcal(\cdot)$ for which updates to $\Lcal(t-1)$ consist only of (1) the deactivation of one leaf denoted by $\ell(t)$ (2) the introduction and activation of the children of $\ell(t)$ in the tree underlying $\Lcal(\cdot)$. Therefore, an update is always of the form,
$$\Lcal(t) = (\Lcal(t-1) \setminus \{\ell(t)\})\cup C_{\ell(t)}.$$
The case where $C_{\ell(t)} = \emptyset$ shall be called a leaf deletion, whereas the case where $C_{\ell(t)} \neq \emptyset$ shall be called a leaf fork. 
More complex updates can be recovered by iteratively applying several of these elementary updates. For any $u\in \Vcal$, we define by $c_u^+(t)$ the number of children of $u$ that have at least one descendant in $\Lcal(t)$, i.e. $c_u^+(t) = \card{\{c \in C_u : \Lcal_c(t)\neq \emptyset\}}$. We then denote by $\bdelta(t)$ the following configuration, 
\begin{equation*}
    \forall u \in \Tcal(t)  :~  \delta_u(t) =  \frac{1}{\prod\limits_{v\in u\rightarrow r}c_{p(v)}^+(t)},\quad \quad \text{and} \quad \quad \forall u\in \Vcal\setminus \Tcal(t):~  \delta_u = 0
\end{equation*}
where $u\rightarrow r$ is the list of ancestors of $u$, including $u$ and excluding $r$. 
One can directly check that $\bdelta(t) \in \Xcal(t)$ by induction on the tree structure (note that $\delta_r(t) =1$ by empty product).\footnote{For $\ell \in \Lcal(t) $,  $\delta_\ell(t)$ can be interpreted as the probability that a random depth-first search (i.e., a DFS initialized at the root that prioritizes neighboring branches at random) hits $\ell$ before any other node of $\Lcal(t)$. We note that the configuration $\bdelta(t)$ also explicitly appears, for a different purpose, in \cite{bubeck2022shortest}.} 
Noticing that $\sum_{v\in \ell\rightarrow r}(c_{p(v)}^+(t)-1)\leq w-1$ and applying Lemma \ref{lem:helper} in the Appendix, we have that $\forall \ell\in \Lcal(t): \delta_\ell(t) \geq 2^{1-w}$.

\paragraph{Proof structure.} 
Let $\bx(\cdot)$ be the algorithm with $(a,b)$ overhead \eqref{eq:assumption}. We shall define an algorithm $\bz(\cdot)$ that satisfies the following two properties on all instances $\Lcal(\cdot)$: 
\begin{equation}\label{eq:bound2}
    \Cost(\bz(\cdot),\Lcal(\cdot))\leq 8\Cost(\bx(\cdot),\Lcal(\cdot)){}+3D,
\end{equation}
\begin{equation}\label{eq:equilibrium1}
    \forall t\in \Nbb, \quad \forall \ell\in \Lcal(t) : z_\ell(t) \leq (2x_\ell(t)-\delta_\ell(t))^+.
\end{equation}
We now explain why these two properties are sufficient to argue about the existence of a $(16b{}+3)$-competitive algorithm for tree traversal. Using $\delta_\ell(t)\geq 2^{1-w}$ and \eqref{eq:equilibrium1}, we have $z_\ell(t) >0 \implies x_\ell(t) \geq 2^{-w}$. Therefore, the cost of $\bx(\cdot)$ must thus satisfy
$$2^{-w}L_z\leq \Cost(\bx(\cdot),\Lcal(\cdot))\leq aL+bD.$$
where $L_z$ is the number of nodes of $\Lcal(\cdot)$ that are attained with non-zero probability by $\bz(\cdot)$. Without loss of generality, one can assume that no leaf $\ell$ is extended or forked at time $t$ if $z_\ell(t) =0$.\footnote{This is because the designer of $\Lcal(\cdot)$ is playing against $\bz(\cdot)$ and has no reason to make a move that diminishes its future options, while inducing no cost to the player. In game-theoretic language, we say that such a strategy is dominated.} Therefore, $L \leq wL_z$.\footnote{Because for any node $u$ that receives $z_u>0$, at most $w-1$ children $v$ with $z_v=0$ may be created.} Using $a<2^{-(w+1)}/w$, we obtain 
$2^{-w}L_z\leq \Cost(\bx(\cdot),\Lcal(\cdot))\leq 2^{-(w+1)}L_z+bD$ and thus, $2^{-(w+1)}L_z\leq bD$. This implies that 
$\Cost(\bx(\cdot),\Lcal(\cdot))\leq 2bD$.
Equation~\eqref{eq:bound2} then allows to conclude that  
\begin{equation*}
    \Cost(\bz(\cdot),\Lcal(\cdot))\leq (16b{}+3)D.
\end{equation*}
We thus now turn our attention to presenting a fractional algorithm $\bz(\cdot)$ satisfying Equations \eqref{eq:bound2} and \eqref{eq:equilibrium1}.

\paragraph{Potential and algorithm definition.} 
For three configurations $\bdelta,\bx,\bz \in \Xcal$,  we define the potential,
\begin{equation}
    D(\bdelta,\bx,\bz) = \sum_{u\in \Vcal}(z_u+\delta_u-2x_u)^+ = \OT^\uparrow(\bz, 2\bx-\bdelta).
\end{equation}
This potential penalizes regions of the tree that violate Equation \eqref{eq:equilibrium1}. Our algorithm $\bz(\cdot)$ is defined at time $t$ as follows, 
\begin{equation}\bz(t) \in \argmin_{\bz\in \Xcal(t)} \OT^\uparrow(\bz(t-1),\bz)+D(t,\bz),\label{eq:alg_potential}
\end{equation}
where ties are broken in favor of a configuration maximizing $\OT^\uparrow(\bz(t-1),\bz)$, and where $D(t, \bz)$ is a shorthand for $D(\bdelta(t),\bx(t),\bz)$. This algorithm is indeed a function of $\Lcal(\leq t)$ because $\bdelta(t)$ and $\bx(t)$ are well-defined functions of $\Lcal(\leq t)$, and the optimizer is over $\Xcal(t)$, so the terms for $u\in \Vcal\setminus \Tcal(t)$ in $D(t,\bz)$ are all equal to $0$.
We now prove that $\bz(\cdot)$ satisfies both desired Equations~\eqref{eq:equilibrium1} and \eqref{eq:bound2}.

\paragraph{Proof of \eqref{eq:equilibrium1}.}
We first observe that our algorithm satisfies at all times the following condition,
\begin{equation}\label{eq:stability}
\forall \bz'\in \Xcal(t): D(t,\bz(t))\leq  \OT^\uparrow(\bz(t),\bz')+D(t,\bz'),
\end{equation}
where the inequality in \eqref{eq:stability} is strict if $\bz' \neq \bz(t)$. 
Indeed, \eqref{eq:stability} is just the combination of the optimality of $\bz(t)$ in \eqref{eq:alg_potential}, $\OT^\uparrow(\bz(t-1),\bz(t))+D(t,\bz(t))\leq \OT^\uparrow(\bz(t-1),\bz')+D(t,\bz')$ and the triangle inequality, $\OT^\uparrow(\bz(t-1),\bz')\leq \OT^\uparrow(\bz(t-1),\bz(t))+\OT^\uparrow(\bz(t),\bz')$.
Further note that in the case of an equality in \eqref{eq:stability}, we must have an equality in both preceding inequalities. By the first equality, we obtain that $\bz'$ is also a minimizer of \eqref{eq:alg_potential}, which, by the tie-breaking rule of \eqref{eq:alg_potential} implies that $\OT^\uparrow(\bz(t-1),\bz')\leq \OT^\uparrow(\bz(t-1),\bz(t))$, the second equality then yields $\OT^\uparrow(\bz(t),\bz') = 0$ and consequently $\bz(t)=\bz'$.

We now prove Equation \eqref{eq:equilibrium1}.
Consider by contradiction a leaf $\ell$ such that $z_\ell(t)> (2x_\ell(t)-\delta_\ell(t))^+$. Denote by $A\in\Vcal$ the first ancestor of $\ell$ such that $z_A(t)\leq (2x_A(t)-\delta_A(t))$. The existence of $A$ is guaranteed by the fact that the root satisfies this condition.
For any node $u\in \ell\rightarrow A$ (i.e. $u$ is an ancestor of $\ell$ and a strict descendant of $A$), we have $z_u(t)>(2x_u(t)-\delta_u(t))$. There must exist a leaf $\ell'$ that is a descendant of $A$ such that for any $u\in \ell'\rightarrow A : z_u(t)<(2x_u(t)-\delta_u(t))$. Consider the configuration $\bz'$ where an arbitrarily small mass $\epsilon$ is moved from $\ell$ to $\ell'$ in $\bz(t)$. We have $\OT^\uparrow(\bz(t),\bz') = \epsilon d(\ell,A)$. Also observe that $D(t,\bz') = D(t,\bz)-\epsilon d(\ell,A)$. This forms a contradiction of the strict version of Equation \eqref{eq:stability} for $\bz'\neq \bz(t)$.

\paragraph{Strategy for proving \eqref{eq:bound2}.} The  cost of a strategy $\bz(\cdot)$ on an instance $\Lcal(\cdot)$ can be decomposed as a sum of two terms,
$$\Cost(\bz(\cdot),\Lcal(\cdot)) = \Cost^\uparrow(\bz(\cdot),\Lcal(\cdot)) + \Cost^\downarrow(\bz(\cdot),\Lcal(\cdot)),$$
where $\Cost^\uparrow(\bz(\cdot),\Lcal(\cdot)) = \sum_{t}\OT^\uparrow(\bz(t-1),\bz(t))$ and  $\Cost^\downarrow(\bz(\cdot),\Lcal(\cdot)) = \sum_{t}\OT^\downarrow(\bz(t-1),\bz(t))$.
At the end of the game, the average depth of the distribution underlying the ultimate configuration is equal to $\Cost^\downarrow(\bz(\cdot),\Lcal(\cdot))-\Cost^\uparrow(\bz(\cdot),\Lcal(\cdot))\leq D$, therefore, 
$$\Cost(\bz(\cdot),\Lcal(\cdot))\leq 2\Cost^\uparrow(\bz(\cdot),\Lcal(\cdot))+D.$$
Thus it remains to bound the quantity $\Cost^\uparrow(\bz(\cdot),\Lcal(\cdot)).$ In what follows, we shall show that 
the following equation holds at all times $t$, 
\begin{align}
    \OT^\uparrow(\bz(t-1),\bz(t))+D(t,&\bz(t))- D(t-1,\bz(t-1))\leq\nonumber\\
    &2\OT(\bx(t-1),\bx(t))+z_{\ell(t)}(t-1)+\sum_{u\in \Vcal}\delta_u(t)-\delta_u(t-1),\label{eq:bound4}
\end{align}
where $\ell(t)$ is the leaf that is forked or deleted at time $t$. Taking the telescopic sums over $t$, and observing that the potential is always non-negative and initially equal to $0$, and that at all times $\sum_{u\mathrel{\ne}r}\delta_u(t) \leq D$ because $\bdelta(t)\in \Xcal(t)$, and that $\sum_{t}z_{\ell(t)}(t-1) \leq 2\Cost(\bx(\cdot),\Lcal(\cdot))$ by \eqref{eq:equilibrium1}, we obtain, 
\begin{align*}
\Cost^\uparrow(\bz(\cdot),\Lcal(\cdot))
&\leq 4\Cost(\bx(\cdot),\Lcal(\cdot))+D.
\end{align*}
Thus, $\Cost(\bz(\cdot),\Lcal(\cdot))\leq 8\Cost(\bx(\cdot),\Lcal(\cdot))+3D$ providing Equation \eqref{eq:bound2} as desired. To complete the proof, we therefore need to prove \eqref{eq:bound4}. 

Note that equations of the type of \eqref{eq:bound4} are central in online algorithms, where they are referred to as potential-based arguments \cite{koutsoupias2009k}, allowing to compare the performance of two algorithms. The unconventional aspect of the present potential (which makes the rest of the proof quite technical) is that it is time-varying, much like the potential in \cite{bubeck2022shortest}.  

\paragraph{Proof of \eqref{eq:bound4} in leaf deletions.}
We first consider the case of the deletion of a leaf $\ell(t)$ at time $t$ (recall that this means that $\ell(t)$ has no descendant in $\Lcal(t)$). We consider $\bar \bz$, the solution of the following optimization problem
\begin{equation}\label{eq:relaxation}
    \bar \bz  = \argmin_{\bz\in \Xcal(t-1)} \OT^\uparrow(\bz(t-1),\bz)+D(\bdelta(t),\bx(t),\bz),
\end{equation}
which is a relaxation of \eqref{eq:alg_potential} because $\Xcal(t)\subset \Xcal(t-1)$ for leaf deletions. By the same reasoning as above, we can show that $\forall \ell \in \Lcal(t-1): \bar z_\ell\leq (2x_\ell(t)-\delta_\ell(t))^+$. In particular, since $x_{\ell(t)}(t) = 0$ and $\delta_{\ell(t)}(t)=0$, we have that $\bar z_{\ell(t)} = 0$, which implies $\bar \bz \in \Xcal(t)$. Since \eqref{eq:relaxation} has the same objective as \eqref{eq:alg_potential}, we have that $\bz(t)$ is also an optimizer of \eqref{eq:relaxation}, which implies, 
\begin{equation}\label{eq:optimdel}
    \OT^\uparrow(\bz(t-1),\bz(t))+D(t,\bz(t))\leq D(t,\bz(t-1)).
\end{equation}
We can then write, 
\begin{align*}
    D(t,\bz(t-1)) &-D(t-1,\bz(t-1)) \\
    &=\underbrace{D(\bdelta(t),\bx(t),\bz(t-1))-D(\bdelta(t-1),\bx(t),\bz(t-1))}_\mathfrak{A}\\
    &+\underbrace{D(\bdelta(t-1),\bx(t),\bz(t-1))-D(\bdelta(t-1),\bx(t-1),\bz(t-1))}_\mathfrak{B}.
\end{align*}
We denote by $K(t) = \Tcal(t-1) \setminus \Tcal(t)$ the list of all nodes that were inactivated (killed) at time $t$ (i.e., such that $\ell(t)$ was their unique active descendant at time $t-1$). For $u\in K(t)$ we have $x_u(t)=0$ and $\delta_u(t)=0$. Conversely, for any $u \not\in K(t)$ we have $\delta_u(t)\geq \delta_u(t-1)$. This yields,
\begin{align*}
    \mathfrak{A} &= \sum_{u\in \Vcal}(z_u(t-1)+\delta_u(t)-2x_u(t))^+-(z_u(t-1)+\delta_u(t-1)-2x_u(t))^+\\
     &= -\sum_{u\in K(t)}\delta_u(t-1) + \sum_{u\in \Vcal \setminus  K(t)}(z_u(t-1)+\delta_u(t)-2x_u(t))^+-(z_u(t-1)+\delta_u(t-1)-2x_u(t))^+\\
     &\leq \sum_{u\in \Vcal} \delta_u(t) -\delta_u(t-1).
\end{align*}
Also, using $\forall a,b\in \Rbb: a^+-b^+\leq |a-b|$, we clearly have that $\mathfrak{B} \leq 2\OT(\bx(t-1),\bx(t))$. Invoking Equation~\eqref{eq:optimdel} and the bounds for $\mathfrak{A}$ and $\mathfrak{B}$, we recover as desired Equation \eqref{eq:bound4}, 
\begin{align*}
    \OT^\uparrow(\bz(t-1),\bz(t))+D(t,&\bz(t))- D(t-1,\bz(t-1))\leq 2\OT(\bx(t-1),\bx(t))+\sum_{u\in \Vcal}\delta_u(t)-\delta_u(t-1).
\end{align*}

\paragraph{Proof of \eqref{eq:bound4} in leaf forks.} We now consider the fork of a leaf $\ell(t)$ at time $t$, which is provided with $c_{\ell(t)}\in \Nbb$ children. Recall that we denote by $C_{\ell(t)}$ the set of children of $\ell(t)$ at the moment of the fork. We consider the configuration $\bz'\in \Xcal(t)$ satisfying $\forall u\in \Tcal(t-1): z_u' = z_u(t-1)$ and $\forall u \in C_{\ell(t)}: z_u' = z_\ell(t-1)/c_{\ell(t)}$. By optimality of $\bz(t)$ in \eqref{eq:alg_potential}, we have,
\begin{equation}
    \OT^\uparrow(\bz(t-1),\bz(t))+D(t,\bz(t))\leq D(t,\bz')\label{eq:fork1}
\end{equation}
Also, like we did for deletions, we use $\forall a,b\in \Rbb: a^+-b^+\leq |a-b|$, to obtain 
\begin{equation}\label{eq:fork2}
    D(\bdelta(t),\bx(t),\bz')- D(\bdelta(t),\bx(t-1),\bz')\leq2\OT(\bx(t-1),\bx(t)).
\end{equation}
We finally observe that $\forall u\in \Tcal(t-1): \delta_u(t)=\delta_u(t-1)$, which implies, 
\begin{align}
    D(\bdelta(t),\bx(t-1),\bz') &= D(t-1,\bz(t-1))+ \sum_{u\in C_{\ell(t)}} \left[\delta_u(t)+z_{\ell(t)}(t-1)/c_{\ell(t)}\right]\nonumber\\
    &=D(t-1,\bz(t-1))+z_{\ell(t)}(t-1)+\sum_{u\in \Vcal}\left[\delta_u(t)-\delta_u(t-1)\right].\label{eq:fork3}
\end{align}
Putting Equations \eqref{eq:fork1}, \eqref{eq:fork2}, \eqref{eq:fork3} together, we recover Equation \eqref{eq:bound4}
\begin{align*}
    \OT^\uparrow(\bz(t-1),\bz(t))+D(t,& \bz(t))- D(t-1,\bz(t-1))\leq\nonumber\\
    &2\OT(\bx(t-1),\bx(t))+z_{\ell(t)}(t-1)+\sum_{u\in \Vcal}\delta_u(t)-\delta_u(t-1).\qedhere 
\end{align*}
\end{proof}

\section{Conclusion}
In this work, we presented a distributed algorithm and a new lower bound for asynchronous collective tree exploration. The distributed algorithm uses a deterministic layered graph traversal \cite{coester2026chasing}, and the lower bound leverages the recent lower bound for fractional layered graph traversal \cite{bubeck2023randomized}.  

The careful reader is perhaps tempted to replace the deterministic layered graph traversal algorithm used in the first part of this paper $\ell(\cdot)$ with a randomized layered graph traversal algorithm. Indeed, quite recently, a new algorithm has been discovered \cite{bubeck2022shortest}, with a competitive ratio in $\Theta(w^2)$, which is exponentially better than its deterministic counterpart and matches the lower bound of \cite{bubeck2023randomized}. At first glance, this appears to yield a randomized, distributed collective tree exploration algorithm with even better guarantees. Though the method is correct (it terminates and explores the entire tree), it does not enjoy the randomized guarantees. This is because the construction of the layered graph traversal instance $\Lcal(\cdot)$ introduced in this paper would then depend on the random choices made by the algorithm, resulting in a model of adversary qualified of ``adaptive adversary'' for the layered graph traversal algorithm \cite{ben1994power}. In contrast, the recent randomized algorithm of \cite{bubeck2022shortest} is obtained against the weaker ``oblivious adversary''. In \cite{ben1994power}, the authors show in a very general setting that the power of randomization against an adaptive adversary is severely limited. For this reason, we conjecture that randomization cannot help improve the competitive ratio of collective tree exploration, but that centralized communications can. 
It seems to us that further leveraging the parallel between the \textit{randomized vs. deterministic} dichotomy in online algorithms and the \textit{centralized vs. distributed} dichotomy in collective algorithms is a promising research direction. 

\paragraph*{Acknowledgements.} We thank the reviewers for their helpful feedback. Laurent Massoulié is partially funded by PR[AI]RIE-PSAI (\mbox{ANR-23-IACL-0008}).


\newpage

\appendix

\section{Background on collective exploration}\label{ap:background-cte}

In this section, we recall the standard definitions and properties of online and offline algorithms for synchronous and asynchronous collective tree exploration. We specifically focus on motivating the notions of competitive ratio and regret used throughout the paper.\footnote{In slight abuse of terminology, in this section, we do not count the root in the node of the trees, so that $n$ is equal to the number of edges.}

\subsection{Synchronous collective tree exploration}

In the classical synchronous model (CTE), all $k$ robots start at the root and move in parallel. At every round, each robot may either stay put or traverse one edge. The cost of an algorithm is the number of rounds elapsed before all nodes of the tree have been visited. 

\begin{definition}[Offline benchmark] For a tree $T$, the offline benchmark $\OPT_{\mathrm{sync}}(T)$ is defined as the minimum number of rounds required by an algorithm knowing the whole tree $T$ in advance to traverse all edges of $T$ with $k$ agents starting at the root.
\end{definition}

\begin{lemma}\label{lem:order} For any $k\in \Nbb$ and for any tree $T$ with $n$ nodes and depth $D$,
\begin{align}\label{eq:encadrement}
    \frac{2n}{k} - D &\leq \OPT_{\mathrm{sync}}(T) \leq \frac{2n}{k} + D 
\end{align}
as well as $D\leq \OPT_{\mathrm{sync}}(T)$, thus,  
\begin{align}\label{eq:ratio}
    \OPT_{\mathrm{sync}}(T) = \Theta\left(\frac{n}{k}+D\right).
\end{align}
\end{lemma}
\begin{proof}
We start by explaining the lower bound in \eqref{eq:encadrement}. After all edges have been explored, all agents can return to the root in at most $D$ additional rounds. At that point, every edge has been traversed at least twice, once in each direction. Since at most $k$ edge traversals can occur in one round, this gives
$
k(\OPT_{\mathrm{sync}}(T)+D)\geq 2n,
$
and therefore $\OPT_{\mathrm{sync}}(T)\geq 2n/k-D$.

For the upper bound in \eqref{eq:encadrement}, consider a depth-first-search tour of the tree, of length $2n$. Split this tour into $k$ consecutive segments, each of length at most $\ceil{2n/k}$. Each robot is assigned one segment: it first reaches the starting endpoint of its segment using at most $D$ rounds, and then traverses all edges in the segment. Thus, all nodes are explored within $\ceil{2n/k}+D$ rounds. 

Moreover, $\OPT_{\mathrm{sync}}(T)\geq D$, since some robot must reach a node of depth $D$. Combining the two lower bounds gives $\frac{1}{3}\left(\frac{2n}{k}+D\right)\leq \OPT_{\mathrm{sync}}(T)$ and thus,  $\OPT_{\mathrm{sync}}(T)=\Theta(\frac{n}{k}+D)$.
\end{proof}

\begin{definition}[Online algorithm] An online algorithm for collective tree exploration is one whose actions at some step only depend on the information that has been collected before that step.  
\end{definition}

The worst-case performance of an online algorithm is usually measured by the competitive ratio (multiplicative overhead over the offline optimum) or the regret (additive overhead over the offline optimum). For collective tree exploration, Lemma \ref{lem:order} allows for a simple characterization of both the competitive ratio and the regret (up to constant factors). 

\begin{definition}[Competitive ratio]
An online algorithm is order $c(k)$-competitive if it explores any tree with $n$ nodes and depth $D$ in at most
$
\Ocal(c(k)(n/k+D))
$
synchronous rounds. 
\end{definition}

\begin{definition}[Regret] An online algorithm has $f(k,D)$-regret if it explores any tree with $n$ nodes and depth $D$ in at most
$
2n/k + \Ocal\left(f(k,D)\right)
$
synchronous rounds.
\end{definition}

\begin{remark}[Return-to-root variant]
    Some variants require all robots to return to the root after the exploration is complete. In that case, the benchmark satisfies
$$
\max\{2n/k,2D\}
\leq
\OPT_{\mathrm{sync}}^{\mathrm{return}}(T)
\leq
\ceil{2n/k}+2D.
$$
Thus, the return-to-root and non-return versions have the same offline benchmark up to constant factors. The distinction matters only when one wants to track sharp leading constants for the competitive ratio or the regret.
\end{remark}

\subsection{Asynchronous collective tree exploration}

In asynchronous collective tree exploration, all robots again start at the root of the tree $T$, and are activated one after each other following an infinite sequence of robot moves $\sigma \in [k]^\Nbb$. The cost is the total number of moves performed before all nodes have been visited.

\begin{definition}[Offline benchmark] For a given tree $T$ and activation sequence $\sigma$, the offline benchmark $\OPT_{\mathrm{asyn}}(T,\sigma)$ is the minimum number of moves required by an algorithm that knows both $T$ and $\sigma$ in advance. 
\end{definition}

\begin{lemma}[Offline benchmark] For every tree $T$ with $n$ nodes and depth $D$, and for every activation sequence $\sigma$, this benchmark satisfies
$$
2n-kD\leq \OPT_{\mathrm{asyn}}(T,\sigma)\leq 2n+kD.
$$
\end{lemma}

\begin{proof}
For the lower bound, note that once the tree is explored, all agents could reach the root in $D$ extra moves, and then all $n$ edges would have been traversed at least twice each. For the upper bound, we denote by $m_i$ the number of moves given to robot $i\in [k]$ in the first $2n+kD$ steps and observe that $\sum_{i\in [k]}(m_i-D)^+\geq 2n$. Thus, an offline algorithm can split a depth-first search path of length $2n$ into $k$ segments of length $s_i = (m_i-D)^+$, and each robot $i\in [k]$ is tasked to (i) reach the first endpoint of their segment (ii) cover their segment of length $s_i$.

\end{proof}

\begin{definition}[Regret] An online asynchronous algorithm has $f(k,D)$-regret if it explores every tree with $n$ nodes and depth $D$ in at most
$
2n + \Ocal\left(kf(k,D)\right)
$
moves.
\end{definition}

We also say that an ACTE algorithm is $c(k)$-competitive if it explores every tree with $n$ nodes and depth $D$ in at most
$
\Ocal(c(k)(n+kD))
$
moves, for every activation sequence. This definition is motivated by the synchronous setting:  by Lemma \ref{lem:asynch-synch}, an order $c(k)$-competitive asynchronous algorithm induces an order $c(k)$-competitive synchronous algorithm.

\begin{lemma}[\cite{cosson2024breaking}]\label{lem:asynch-synch}
    An asynchronous exploration algorithm finishing in $f(k,n,D)$ moves for any tree $T$ entails a synchronous exploration algorithm that finishes in $\ceil{f(k,n,D)/k}$ time steps.
\end{lemma}

\begin{proof}[Proof of Lemma \ref{lem:asynch-synch} (dealing with synchronous moves).] The description of ACTE is sequential, as the adversary picks one robot at each $t$. We can relax this assumption by letting the adversary choose an arbitrary subset of robots $I(t)\subset [k]$, which may move at round $t$. This generalized setting encapsulates the synchronous setting for which $\forall t: I(t) = [k]$, but it does not increase the power of the adversary. This is because robots moving away from the same node at the same time may use a tie-breaking rule to emulate sequential decisions before moving simultaneously.\footnote{This assumption is also used in \cite{fraigniaud2006collective}: it is necessary. Otherwise, all such robots would move along the same edge. To deal with robots traversing an edge simultaneously in opposite directions, we utilize the fact that robots do not read or write at their destination.}  
\end{proof}

\begin{remark}[Return-to-root variants]
There is no return-to-root variant in the asynchronous setting because agents may be blocked indefinitely.
\end{remark}

\section{Proof of lemmas}\label{sec:generalization-proofs}

\subsection{Lemmas from Section \ref{sec:setting}}\label{ap:lem-model}

\lemweighted*

\begin{proof}[Proof of Lemma \ref{lem:weighted-edge} (generalizing DACTE to weighted trees)]
    We first assume that all edge weights of the unknown tree $T$ take values in $\Nbb$, i.e., that $a=1$. In this case, the agents can construct the unweighted tree $T'$ (i.e., with edge weights equal to $1$) by subdividing the edges of $T$ with length $\geq 2$ with (imaginary) intermediary nodes. The depth of the unweighted tree $T'$ is equal to the weighted depth $D$ of $T$, and the number of edges $n-1$ of $T'$ is equal to the sum of all edge weights $L$ of $T$. When the adversary selects an agent located at some node $u$ of $T$, this agent shall choose to move to the first node $v$ of $T$ that it would reach if it were exploring $T'$ and if it were provided moves by the adversary indefinitely until it reached a node of $T'\cap T$. The cost of exploring the weighted tree $T$ with this procedure is less than the cost of the unweighted exploration algorithm in $T'$ because all movements that happen in $T$ would have happened in $T'$. 
    The algorithm thus terminates with a cost of at most $f(k,L,D)$. 
    We then assume that all edge weights are in $a\Nbb$ for some known $a>0$. The agent can apply the above algorithm to the tree (of integral edge lengths) obtained by multiplying all edge weights by $1/a$. After rescaling, the resulting cost is therefore in $af(k,L/a,D/a)$. This is equal to $f(k,L,D)$ if $f(\cdot,\cdot,\cdot)$ is linear in its second and third arguments.
\end{proof}

\lemcontinuous*

\begin{proof}[Proof of Lemma \ref{lem:continuous} (generalizing DACTE to continuously moving agents)]
    A robot moving inside an edge between two nodes attains its destination at an instant that depends on the speed provided by the adversary. For $i\in [k]$ and $j\in \Nbb$, we denote by $\tau^{(i)}_j\in \Rbb^+$ the instant at which the $i$-th robot leaves a node\footnote{This matches the definition of \textit{visiting} a node} for the $j$-th time. At $\tau^{(i)}_j$, the $i$-th robot must read/write from the whiteboard on which it is located as well as select an adjacent edge to traverse next. This is done instantaneously\footnote{The asynchrony we study is on the speeds of the robots, not on concurrent memory access.} with the discrete-time algorithm. We can assume without loss of generality that no two values of $(\tau^{(i)}_j)_{i\in [k], j\in \Nbb}$ are equal because robots can communicate and coordinate when they are located on the same node at the same instant.  We now consider the values of $(\tau^{(i)}_j)_{i\in [k], j\in \Nbb}$ in increasing order, which allows to define a sequence of indices $i(1), i(2),\dots$, such that for $t\in \Nbb$, $i(t)$ is the index of the robot that leaves a node for the $t$-th time. We note that the movements in the continuous settings will be the same as those in the discrete setting, where the adversary selects indices $i(1), i(2), \dots$. This observation crucially relies on the fact that, in the sequential definition of the problem (Section~\ref{sec:model}), a robot does not write at its destination. 
\end{proof}

\proptargets*
\begin{proof}[Proof of Proposition \ref{prop:targets}] Consider the potential $P(t) = \sum_{i\in [k]} d(p_i(t),v_i(t))$, where $p_i(t)$ is the position of robot $i$ at time $t$. Observe that the potential increases by $1$ at any move $t$ where the team discovers a new edge \eqref{it: 1}, decreases by $1$ at any move $t$ where a robot moves towards its target \eqref{it: 2} and increases by at most $d(v_i(t),v_i(t+1))$ when the robot $i$ changes target. Also recall that at all times $t$, $\one(t: \eqref{it: 1})+\one(t: \eqref{it: 2})=1$. We bound the variation of the potential as follows:
\begin{align*}
    P(t+1)-P(t) &\leq \one(t: \eqref{it: 1})- \one(t: \eqref{it: 2}) +\sum_{i\in k}d(v_i(t),v_i(t+1))\\
    &= 2\one(t: \eqref{it: 1})-1+\sum_{i\in k}d(v_i(t),v_i(t+1)).
\end{align*}
We then recover the announced result by a telescopic sum, using $0\leq P(M) -P(0)$.
\end{proof}

\subsection{Lemmas from Section \ref{sec:lower-bound}}\label{ap:lem-lb}

\lemacte*
\begin{proof}[Proof of Lemma \ref{lem:acte-ltt} (showing that an ACTE algorithm induces a tree traversal algorithm)] Let $\Lcal(\cdot)$ be some tree traversal instance on which we shall define our fractional algorithm $\bx(\cdot)$. The idea is that a team of $k$ robots will be using the ACTE algorithm in the tree underlying the instance, $T = \cup_t \Tcal(t)$. Note that the number of nodes of $T$ is denoted by $n=L$ and its depth by $D$. The algorithm $\bx(\cdot)$ is defined by induction. At any step $t$ of the tree traversal, the fractional configuration $\bx(t)$ shall represent the distribution of $k$ robots in $T$, where all robots are located on $\Lcal(t)$, each robot being represented by a fractional mass of $1/k$. Then, when the layer $\Lcal(t+1)$ is revealed to the tree traversal algorithm, 
we consider an asynchronous adversary of the ACTE algorithm that lets robots move in $T$ until they all hit $\Lcal(t+1)$ but no further, and at this point, their distribution defines $\bx(t+1)\in \Xcal(t+1)$. Note that $\bx(t+1)$ is indeed a function of $\Lcal(\leq t+1)$, because no robot has had the chance to gain information outside $\Tcal(\leq t+1)$. The cost of the tree traversal algorithm in step $t$, which equals $\OT(\bx(t),\bx(t+1))$, is less than the cost of the ACTE algorithm (counting the number of moves of the robots), divided by a factor $k$. Overall, this yields $\Cost(\bx(\cdot),\Lcal(\cdot)) \leq \frac{c(k)}{k}(n+kD) = \frac{c(k)}{k}L +c(k) D$. 
\end{proof}

\begin{lemma}\label{lem:helper}
Let $w\in \Nbb$. For any $(c_i)_{i\in \Nbb} \in \Nbb^\Nbb$ satisfying $\sum_{i\in \Nbb}(c_i-1)\leq w$, we have $\prod_{i\in \Nbb} c_i \leq 2^w$. 
\end{lemma}
\begin{proof}
We aim to compute the value of the optimization problem of maximizing $\prod_{i\in \Nbb} c_i$ for integral variables $(c_i)_{i\in\Nbb}\in \Nbb^\Nbb$ subject to the constraint $\sum_{i\in \Nbb}(c_i-1)\leq w$. We first notice that there is a solution to this problem for which all variables take values in $\{1,2\}$, and conclude by observing that the maximum is attained when the largest amount of $2$ is selected, i.e., when the value is $2^{w}$. Consider two integers, $c_1, c_2 \in \Nbb$ satisfying $c_1\geq c_2+1$. We have $(c_1-1)(c_2+1) = c_1c_2+c_1-c_2-1 \geq c_1c_2$. Therefore, from any $(c_i)_{i\in \Nbb}$, if there exists $i$ such that $c_i\geq 3$, then it is possible to decrease this value by $1$ and increase the value of some other $1$ to $2$, while not decreasing the value of the optimum.     
\end{proof}

\section{Failure of known algorithms for DACTE}\label{sec:failure}
In this section, we explain why previous algorithms fail to provide non-trivial competitive guarantees for distributed asynchronous collective tree exploration (DACTE). 
\paragraph*{Greedy algorithms.} The class of greedy exploration algorithms, defined by \cite{higashikawa2014online}, includes the synchronous $\Ocal(k/\log k)$-competitive algorithm of \cite{fraigniaud2006collective} as its main example. Following \cite{higashikawa2014online}, for a node $u\in T$, we say that the subtree rooted at $u$ is \textit{finished} if all edges below $u$ have been explored and
there is no robot (strictly) below $u$. An exploration algorithm is \textit{greedy} if the move of a robot is directed towards the root only if it is located atop a finished subtree. The algorithm in \cite{fraigniaud2006collective} consists of splitting the robots at a node evenly between the unfinished subtrees dangling below their position. We observe that greedy algorithms cannot achieve a finite time guarantee in the asynchronous setting (DACTE). Consider, for instance, the case where two robots are co-located at the top of a finished subtree. If the adversary provides a (possibly infinite) series of moves to one of the robots but not to the other robot, then the state of the exploration does not progress because the first robot cannot leave the other. It is tempting to try to bypass this issue by slightly reducing the power of the adversary, e.g., by forcing the adversary to allow co-located robots to move synchronously, or by relaxing the definition of a finished subtree.\footnote{One could say, e.g., that a subtree is finished as soon as all of its edges have been explored. Note that with distributed communications, it is unclear how an agent would then know whether it is at the top of a finished subtree.} Even then, it is easy to see that on the ``comb'' tree of \cite{higashikawa2014online}, for which $n = \Theta(D^2)$, any asynchronous greedy algorithm will require $kn/2$ robot moves, because consecutive branches can be forced to be explored one after another, each with a team of at least $k/2$ robots. There is thus no hope of obtaining guarantees of the form $2n+f(k)D$, or of the form $c(k)(2n/k +D)$ with $c(k) = o(k)$ for DACTE, using greedy algorithms.

\paragraph*{Depth-first search.} The simplest algorithm (and only algorithm to date) to achieve linear guarantees for DACTE is a variant of depth-first search. More precisely, consider the algorithm in which all robots are following the trajectory of a \textit{leader}, which is the robot that has been given the most moves by the adversary so far. Since the robots are on the same trail, they can tell whether they are the leader or a \textit{follower} (even under distributed communications). If they are the leader, they adopt a depth-first search approach: they go through an unexplored edge if one is adjacent, and proceed towards the root otherwise. If they are not the leader, they adopt a follower attitude, following the leader's trail. This algorithm finishes in at most $\Ocal(2kn)$ total moves, since each robot goes through each edge at most twice (by property of depth-first search). Clearly, this algorithm does not achieve a linear-in-$D$ regret nor an $o(k)$-competitive ratio. 

\paragraph{Breadth-first depth-next.} We note that the ``breadth-first depth-next'' algorithm of \cite{cosson2023efficient} with a regret in $\Ocal(\log(k) D^2)$ can handle an asynchronous adversary. However, the regret guarantee is not linear in $ D$ and thus fails to provide a finite competitive ratio. Another reason why linear algorithms (in $n$ and $D$) are desirable is that they can be used on weighted trees (unlike algorithms with super-linear guarantees) because they are scale-invariant (see Section~\ref{sec:extensions}). Additionally, while most work on collective exploration has focused on obtaining worst-case guarantees, quantified by the competitive ratio or regret, it is natural to wonder about average case guarantees \cite{roughgarden2021beyond}, i.e., to obtain algorithms with good performance on trees sampled uniformly at random. There has been a lot of research on trees sampled uniformly at random (see e.g. the pioneering work of \cite{aldous1993continuum}), and it is well established that their size $n$ scales quadratically with their depth $D$, with high probability. Therefore, a linear regret algorithm for DACTE yields an asymptotically optimal guarantee in the average case setting, because the regret is asymptotically dominated as $n$ scales to infinity.

\section{Distributed algorithm}

\subsection{Formal description}\label{sec:formal-desc}
We now provide the formal description of our algorithm in the distributed communication model.  
The algorithm uses registers available both on the whiteboards and inside the robot's internal memory. Initially, all registers are empty. We start by defining the registers of the whiteboards. 
\begin{itemize}
    \item If the node $u\in T$ was already visited by some robot, the following registers are defined,
    \begin{itemize}
        \item $F_u\subset E$: the list of all unexplored edges below $u$ ;
        \item $C_u^+\subset C_u$: the list of children $c$ of $u$ such that the edge $(u,c)$ was explored by a robot which has not revisited $u$ since ;
    \end{itemize}
    \item If $u$ was elected as a target by some robot that has visited $u$, the following registers are additionally defined,
    \begin{itemize}
        \item $h_u\in \Nbb$: the current index of this node in the list of targets, since $u$ may appear multiple consecutive times in the list of target, $h_u$ will possibly be incremented in the execution of the algorithm; 
        \item $\Lcal_u(\leq h_u)$: an instance of tree traversal of length $h_u$ ;
    \end{itemize}
    \item If condition \eqref{it: change_anchor} was raised by some robot at $u$, the following registers are additionally defined (they will never be updated), 
    \begin{itemize}
        \item $v^{\text{next}}_u$: the target succeeding $u$ ;
        \item $\Lcal^{\text{next}}_u$: the layer succeeding $\Lcal_u(h_u)$.
    \end{itemize}
\end{itemize}

\noindent
We then define the registers available in each robot's memory, indexed by $i\in [k]$:
\begin{itemize}
    \item $p_i\in \Vcal$: its current position ;
    \item $v_i\in \Vcal$: its current target ;
    \item $c_i\in \Vcal$: the last children of a target that robot $i$ has visited. In other terms, the last node $u\in T$ that was visited by $i$ and for which $h_u$ was not defined; 
    \item $h_i\in \Nbb$: the index of the robot's target ;
    \item $\Lcal_i(\leq h_i)$: an instance of tree traversal of length $h_i$. 
\end{itemize}
The registers are read and updated by the agents via the exploration algorithm. We specify all such changes, except for those to $F_u, C_u$, and $p_i, c_i$, which are obvious.

We now describe the exploration algorithm from the point of view of a robot $i$ which is allowed to move by the adversary, i.e., at some $t\in\Nbb$ for which $i(t)=i$. Depending on the robot's situation, the primitive used by the robot varies. The choice of primitive depends on three simple questions that the agent can answer with local information (1) is the agent adjacent to an unexplored edge? (2) is the agent located at its target? (3) is $v^{\text{next}}_u$ defined at this node? It is a direct observation that the agent has no trouble choosing the correct routine. We list all possible cases below.

\begin{itemize}
    \item \textbf{Locally-Greedy.} If robot $i$ is adjacent to some unexplored edge or is not located at its target, then it simply applies \eqref{it: 1} and \eqref{it: 2} with procedure \texttt{Locally-Greedy}$(i)$. This corresponds to the case where the condition \eqref{it: change_anchor} is not raised by the robot ;
    \item \textbf{Follower.} If robot $i$ raises condition \eqref{it: change_anchor} at its target $u =v_i$, and $v^\text{next}_u$ is defined, then robot $i$ updates its target to $v^\text{next}_u$ and moves towards it \eqref{it: 2} ;
    \item \textbf{Leader.} Finally, if robot $i$ raises condition \eqref{it: change_anchor} at its target $u=v_i$ and $v^\text{next}_u$ is undefined, then it runs the procedure \texttt{Leader}$(i)$. The role of this procedure is to define $v^\text{next}_u$ and to write it at $u$ so that consecutive robots will follow the trail of the leader. Importantly, since some children of $u$ might still have unexplored descendants, the set $C_u^+$ of all such children is added to the active leaves of the layered tree traversal instance. 
\end{itemize}
Also, at the beginning of any move in which the robot $i$ is located on its target $u$ and $v^{\text{next}}_u$ is undefined, the robot calls the routine \texttt{Synchronize}. The routine is responsible for updating the node registers $h_u$ and $\Lcal_u(\leq h_u)$. Note that these registers are no longer modified once $v^{\text{next}}_u$ is defined.
In Algorithms~\ref*{alg:follower}--\ref*{alg:synchronize}, copying a layer between a robot and a node also copies all preceding layers from the source into the corresponding entries at the destination.
Initially, the registers at the nodes are all empty, and those of robots are all defined by $p_i \gets r ; \quad v_i \gets r ; \quad h_i\gets 1; \quad \Lcal_{i}(1) \gets \{r\}$.

\begin{algorithm}[H]
\caption{\texttt{Locally-Greedy}}\label{alg:locally-greedy}
\begin{algorithmic}[1]
\Require The robot $i$ is adjacent to an unexplored edge or is not located at its target $v_i$.
\If{robot is adjacent to some unexplored edge $e$} \Comment{Apply rule \eqref{it: 1}.}
\State $\texttt{Move-Along}(e)$ 
\Else\Comment{Apply rule \eqref{it: 2}.}
\State $\texttt{Move-Towards}(v_i)$ 
\EndIf
\end{algorithmic}
\end{algorithm}

\vspace{-1em} 

\begin{algorithm}[H]
\caption{\texttt{Follower}}\label{alg:follower}
\begin{algorithmic}[1]
\Require The robot $i$ is not adjacent to an unexplored edge and is located at its target $u=v_i$ for which $v^{\text{next}}_u$ is defined.
\State $h_i\gets h_u+1 ;  \quad  \Lcal_i(h_i)\gets\Lcal^{\text{next}}_u ; \quad v_i\gets v^{\text{next}}_u$
\State \texttt{Move-Towards}($v_i$) \Comment{Apply rule \eqref{it: 2}.}
\end{algorithmic}
\end{algorithm}

\vspace{-1em} 

\begin{algorithm}[H]
\caption{\texttt{Leader}}\label{alg:leader}
\begin{algorithmic}[1]
\Require The robot $i$ is not adjacent to an unexplored edge and is located at its target, $u=v_i$ for which $v^{\text{next}}_u$ is undefined. 
\State $\Lcal^{\text{next}}_u\gets \left(\Lcal_u(h_u)\setminus\{u\}\right) \cup C_u^+$
\If{$\Lcal^{\text{next}}_u \neq \emptyset$}\Comment{Exploration is finished otherwise!}
\State $h_i\gets h_u+1 ; \quad \Lcal_i(h_i)\gets \Lcal^{\text{next}}_u$
\State $v_i\gets v^{\text{next}}_u\gets \ell(\Lcal_i(\leq h_i))$
\Comment{Where $\ell(\cdot)$ denotes a lazy deterministic tree traversal algo.}
\State \texttt{Move-Towards}($v_i$) \Comment{Apply rule \eqref{it: 2}.}
\EndIf
\end{algorithmic}
\end{algorithm}

\vspace{-1em} 

\begin{algorithm}[H]
\caption{\texttt{Synchronize}}\label{alg:synchronize}
\begin{algorithmic}[1]
\Require The robot $i$ is located on its target $u$ and $v^{\text{next}}_u$ is not defined. 
\If{$u$ is visited as a target for the first time}\Comment{Initialization of the registers.}
\State $h_u\gets h_i ; \quad \Lcal_u(h_u) \gets \Lcal_i(h_i)$
\ElsIf{$u$ is visited as the target of robot $i$ for the first time}
\State $h_u\gets h_u+1$
\State $\Lcal_u(h_u)\gets \Lcal_u(h_u-1)\setminus \{c_i\}$
\EndIf
\end{algorithmic}
\end{algorithm}

\begin{figure}
    \centering
\begin{tikzpicture}[
  grow=down,
  level distance=10mm,
  level 1/.style={sibling distance=28mm},
  level 2/.style={sibling distance=22mm},
  level 3/.style={sibling distance=18mm},
  level 4/.style={sibling distance=16mm},
  dot/.style={circle,draw,minimum size=6pt,inner sep=1.5pt},
  solid/.style={fill=black},
  hollow/.style={fill=white},
  doubleb/.style={
  double,                
  double distance=0.8pt,
  preaction={             
      draw,
      double,
      double distance=1.8pt 
    }
  },
  dedge/.style={edge from parent/.style={draw,densely dashed}},
  edge from parent/.style={draw},
  every label/.style={font=\scriptsize,inner sep=0.5pt}
]

\node[dot,solid,label=45:$v_0$] (v0) {} 
  child[xshift=-10mm] { node[dot,hollow] {} }
  child { node[dot,solid,label=45:$v_1$] (v1) {}
    child[xshift=-8mm] { node[dot,solid,doubleb,label=45:$v_2$] {}
      child[xshift=-6mm] { node[dot,doubleb] {}
        child[xshift=-6mm] { node[dot,hollow] {} }
        child[dedge,xshift=-10mm] { node[coordinate] {} }
        child[dedge,xshift=-10mm] { node[coordinate] {} } } }
    child { node[dot,solid,label=45:$v_3$] (v2) {}
      child[xshift=-8mm] { node[dot,hollow] {}
        child[xshift=-6mm] { node[dot,hollow] {} } }
      child { node[dot,solid,label=45:$v_4$] (v3) {}
      } } }
  child[xshift=35mm] { node[dot,solid,doubleb,label=45:$v_5$] (v4) {}
    child { node[dot,doubleb] {}
      child[dedge] { node[coordinate] {} } }
    child { node[dot,hollow] {}
      child { node[dot,hollow] {} } }
    child { node[dot,doubleb] {}
      child[dedge,xshift=-3mm] { node[coordinate] {} }
      child[dedge,xshift= 3mm] { node[coordinate] {} } }
    child[dedge,xshift=3mm] { node[coordinate] {} }
  };

\end{tikzpicture}
    \caption{Illustration of the algorithm, with $k=4$. The tree represents all discovered nodes and edges at a given instant. Discovered but unexplored edges appear as dashed. Black-filled nodes are nodes that have been elected as targets in the past, and double black nodes $\{v_2,v_5\}$ are current targets. Double hollow nodes are nodes that will potentially become targets. Single hollow nodes are nodes that will never become targets, because the exploration of their subtree ended before the traversal algorithm selected them. The $i$-th node to be selected by the layered graph traversal algorithm is denoted by $ v_i$. This index does not correspond to the order in which the nodes were discovered, but instead to the order in which they were elected as targets by a robot. The consecutive layers in this execution of the algorithm is $\Lcal(0) = \{v_0\}$, $\Lcal(1) = \{v_1,v_5\}$, $\Lcal(2) = \{v_2,v_3,v_5\}$, $\Lcal(3) = \{v_2,v_3,v_5\}$, $\Lcal(4)= \{v_2,v_4,v_5\}$, $\Lcal(5) = \{v_2,v_5\}$ (other sequences would be compatible with the Figure). At the instant considered, there must be one robot exploring below $v_2$ and at least two robots exploring below $v_5$.  The last robot could be transferring between a previous target (such as $v_3$ or $v_4$) and $v_5$. Note that, due to distributed communications, no robot has a full picture of the explored tree.}
    \label{fig:illustration-algo}
\end{figure}

\subsection{Analysis}\label{sec:analysis}

\begin{theorem} The \DACTE algorithm presented in Section \ref{sec:formal-desc} explores any tree of $n$ nodes and depth $D$ with $k$ robots in at most
$2n + kc_k D$ moves, where $c_k$ denotes the competitive ratio of the width-$k$ layered graph traversal algorithm $\ell(\cdot)$ used as a subroutine. 
\end{theorem}

\begin{proof} We make a series of claims that together provide the proof of the statement. 

\paragraph*{Claim 1.} The algorithm described in Section \ref{sec:formal-desc} is a locally-greedy algorithm with targets, in the sense of Definition \ref{def:locally-g}.

\begin{proof} It suffices to verify that the robots always respect the rules of locally-greedy algorithms with targets (cf. Definition \ref{def:locally-g}). First, observe that the moving robot is always assigned a move \eqref{it: 1} or \eqref{it: 2} by procedures \texttt{Leader}, \texttt{Follower}, \texttt{Locally-Greedy} except in the case of procedure \texttt{Leader} when $\Lcal_u^{\text{next}}=\emptyset$, but this will imply that exploration is finished. Second, observe that procedure \texttt{Locally-Greedy} is always chosen when the robot is adjacent to an unexplored edge, which enforces the priority of \eqref{it: 1} over \eqref{it: 2}. 
\end{proof}

\paragraph*{Claim 2.} There is a sequence of nodes $S = v^1,v^2, \dots$ such that the sequence of targets chosen by any robot during the exploration is $S$ or a prefix of $S$. Also, at any point of the algorithm, the quantity $v^{\text{next}}_u$ is defined on all targets except for the last one.

\begin{proof} First, observe that for any node $u\in T$ the variable $v^{\text{next}}_u$ is defined once and for all when some robot calls the procedure \texttt{Leader} at $u$. After that call, no changes to any of the registers $h_u,v^{\text{next}}_u,\Lcal_u(\leq h_u),\Lcal^{\text{next}}_u$ will be allowed. Then observe that the target of a robot only changes when that robot calls procedure \texttt{Leader} or \texttt{Follower}, and that procedure \texttt{Follower} at $u$ results in updating the target of the robot to $v^{\text{next}}_u$. Furthermore, calls to \texttt{Leader} or \texttt{Follower} only happen if the robot is located at its target. Therefore, the sequence of targets $S = v^1, v^2, \dots$ is defined by the order in which these nodes were first elected as targets and remains the same for all robots. Note that it remains possible that some robots lag behind and never attain the more advanced targets of the sequence.
\end{proof}

\paragraph*{Claim 3.} All partial sequences $\Lcal_i(\leq h_i)$ stored by the robots and $\Lcal_u(\leq h_u)$ stored on the nodes are consistent in the sense that the longer instances contain the shorter instances as their prefix. Hereafter, we drop the subscripts and denote $\Lcal(\cdot) = \Lcal(1),\Lcal(2),\dots$.   

\begin{proof}
    We start by analyzing all updates made to the instances stored on the whiteboards $\Lcal_u(\leq h_u)$. These changes happen in the procedure \texttt{Synchronize}. We note that such updates necessarily happen at a node $u$ that is a target and such that $v^{\text{next}}_u$ is undefined. Following Claim 2, there is only a single such node $u$ at a given time, and that node is also the most advanced target of the previously defined sequence $S$. Also observe that all updates to $\Lcal_u(\leq h_u)$ 
    consists in either appending one additional element to the list, while not touching previous elements, or, alternatively, of copying a sequence carried by a robot on the register at $u$ (when $u$ is initialized). In the latter situation, the procedure \texttt{Leader} and \texttt{Follower}, combined with Claim 2, guarantees that the instance is a direct copy of the instance at the preceding target, with one additional layer. This maintains the consistency of the layered sequences that are stored on the nodes. The consistency of the sequences stored by the robots follows because they are copies of the sequences stored on the nodes. 
\end{proof}

\paragraph*{Claim 4.} The sequence $\Lcal(\cdot)$ corresponds to a valid layered tree instance in the sense of Section~\ref{sec:inputlgt}. Its underlying tree $\cup_h \Tcal(h)$ is a subtree of $T$, and the aforementioned sequence of targets $S = v^1,v^2,v^3,\dots,$ is the output of the (lazy) layered tree traversal algorithm $\ell(\cdot)$ on $\Lcal(\cdot)$.

\begin{proof}
    The update to $\Lcal(h)$ are either of the form $\Lcal(h+1)\gets \Lcal(h)\setminus \{c_i\}$ or of the form $\Lcal(h+1)\gets \Lcal(h)\setminus \{u\}\cup C_u^+$, where $u\in \Lcal(h)$ and $C_u^+$ contains a subset of children of $u$. This proves that $\Lcal(h+1)$ remains a layer (by induction) and that we have $\forall h: \Lcal(h)\succeq \Lcal(h+1)$. It is obvious that $\Lcal(h)\subset T$, where $T$ is the tree being explored by the team. It is then clear from procedure \texttt{Leader} that $S$ is the output of the lazy layered tree traversal $\ell(\cdot)$ on $\Lcal(\cdot)$. Note that the laziness (defined in Section \ref{sec:lgt}) of $\ell(\cdot)$ is used in the updates of the form $\Lcal(h+1)\gets \Lcal(h)\setminus \{c_i\}$ because the target does not change in these updates. 
\end{proof}

\paragraph{Claim 5.} Whenever $\Lcal(h)$ is first defined, all unexplored edges are below $\Lcal(h)$. The exploration algorithm is correct in the sense that it provides a move to any robot selected by the adversary until all edges have been explored.  
\begin{proof} We will show the statement by induction. The property is true initially, because $\Lcal(0) = \{r\}$. We assume that the property is true for some $h$ and aim to show when $\Lcal(h+1)$ is defined. We consider the two types of updates to $\Lcal(h)$. The first type of update is $\Lcal(h+1)\gets \Lcal(h)\setminus \{c_i\}$. For this type of update, it suffices to justify that there cannot be any unexplored edge below $c_i$ when this update happens. 
Observe that at the moment when robot $i$ left the subtree rooted at $c_i$, it was the only robot to have ever visited the subtree rooted at $c_i$ (because $c_i$ was not a target at that time). Since the algorithm is locally-greedy (Claim 1) robot $i$ must have effectively executed depth-first-search below $c_i$, which means that it did not leave any unexplored edge in the subtree rooted at $c_i$.
We now consider the other type of update $\Lcal(h+1)\gets \Lcal(h)\setminus\{u\} \cup C_u^+$. Note that these updates happen only when all edges adjacent to $u$ have been explored. For the same reason as above, only the children in $C_u^+$ from which no robot has ever returned could lead to unexplored edges. This finishes the induction. To conclude, the algorithm provides a valid move to any robot until some robot observes that $\Lcal(h)=\emptyset$. In this case, there are no more unexplored edges. The algorithm is thus correct.
\end{proof}

\paragraph{Claim 6. } The width of the layered tree instance $\Lcal(\cdot)$ is at most $k$.

\begin{proof}
    We prove this claim by associating a robot to each node $c\in \cup_h \Lcal(h)$ in a way that no two nodes in a same layer are associated to the same robot.\footnote{An exception is made for the root to which we match no robot.} This proves the claim, since there are $k$ robots in total. We consider some node $c\in \cup_h \Lcal(h)$ that is different from the root and denote its parent by $u$. We consider the first moment when $c$ was added to $\Lcal(\cdot)$, in an update of the form  $\Lcal(h_u+1)\gets \Lcal_u(h_u)\setminus \{u\}\cup C_u^+$. Since $c\in C_u^+$ when that update happens, there was a robot below $c$ at this point, and by rule \eqref{it: 1} of locally-greedy algorithms, this robot $i(c)$ is uniquely defined (before the update, there were unexplored edges available at $u$ preventing any other robot from traversing $(u,c)$). Note that $i(c)$ is also defined as the first robot that explored the edge $(u,c)$.  We will now demonstrate that no two nodes within the same layer are associated with the same robot.
    
    We will prove this property by induction on $h$. We can restrict our attention to updates of the form $\Lcal(h_u+1)\gets \Lcal_u(h_u)\setminus\{u\} \cup C_u^+$, since other updates only remove nodes from the layer. We consider one such update at a node $u$, and assume that the property was satisfied at all previous updates. To complete the induction, it suffices to prove that none of the robots associated with nodes in $C_u^+$ are associated with nodes in $\Lcal_u(h_u)\setminus\{u\}$. We thus consider a robot $i$ that is in a subtree below $ c \in C_u^+$ when the update occurs. There are two possible cases: 
    (1) robot $i$ was already in that subtree before $u$ was even elected as its target. 
    In this case, the node $u$ is already associated to $i$ and by the induction hypothesis, robot $i$ not is associated to any other node in $\Lcal_u(h_u)\setminus\{u\}$ 
    (2) robot $i$ attained $u$ after $u$ was elected as its target. In particular, robot $i$ visited all targets of the sequence $S$ until $u$. 
    In this case, consider by contradiction another node $c'\in \Lcal_u(h_u)$ which is associated with $i$ and denote its parent by $u'$. $u'$ is a target preceding $u$ in $S$. Consider the value of $c_i$ right before robot $i$ visited node $u$ and notice that it must be the case that the value of $c_i$ is equal to $c'$. Thus $c_i$ was removed from $\Lcal_u(h_u)$ before $i$ could explore an edge below $u$, which forms a contradiction. Therefore, any layer of $\Lcal(\cdot)$ is of cardinality at most $k$.  
\end{proof}

To conclude, using Claims 1, 4, and 5 and Proposition \ref{prop:targets},  we obtain that the locally-greedy algorithm we defined runs in $T$ for at most $M$ moves, where $M$ satisfies
\begin{equation}\frac{1}{2}\left(M - \sum_{i\in [k]}\sum_{t< M} d(v_i(t),v_i(t+1))\right) \leq |E|=n-1.    \label{eq1}
\end{equation}
The left-hand side of the equation represents a lower bound on the number of edges explored during the algorithm's execution, while the right-hand side denotes the total number of edges to explore. By Claim 4, the tree $T$ is explored when the algorithm stops. Using Claim 2 defining $S = v^1,v^2,\dots$, 
we have that 
\begin{equation}\sum_{i\in [k]}\sum_{t< M} d(v_i(t),v_i(t+1)) \leq k\sum_{h} d(v^h,v^{h+1}).\label{eq2}\end{equation}
Also, using Claim 3, 4, 6 and the definition of deterministic tree traversal algorithm in Section \ref{sec:inputlgt}, 
\begin{equation}\sum_{h} d(v^h,v^{h+1})\leq c_k D,\label{eq3}\end{equation}
where $c_k$ is the competitive ratio associated with the layered graph traversal algorithm $\ell(\cdot)$. The combination of Equations \eqref{eq1}, \eqref{eq2}, and \eqref{eq3} then allows us to conclude that the algorithm explores the underlying tree in at most
\begin{equation*}M \leq 2n+ kc_kD \quad \text{moves.}\qedhere\end{equation*}
\end{proof}

\subsubsection{Relaxation of the communication model}
We observe that the algorithm described in Section \ref{sec:formal-desc} makes limited use of the whiteboards and of the robot's storage capabilities. As an illustration, we briefly describe an alternative (non-classical) distributed communication model with a \textit{unique mobile whiteboard} where our algorithm can be implemented. 

First note that the whiteboards are mainly used at nodes that have been elected as targets. We therefore observe that we need at most $\Ocal(c_k D)$ 
such whiteboards. Interestingly, this quantity is independent of $n$, the number of nodes in the tree. We also note that when a target $u$ has a successor defined $v^{\text{next}}_u$, then all other information at this target becomes useless. Our algorithm can thus be implemented in a model where all nodes are equipped with limited storage capabilities (just enough to store the position of the next target and lists of adjacent explored edges), and the most advanced robot (the leader) moves one large whiteboard to the most advanced target. Finally, we observe that in the absence of such a large whiteboard, one robot could be tasked to play that role while the rest of the $k-1$ robots implement the exploration algorithm.\footnote{This remark is for the synchronous model since it requires that the whiteboard-robot is always able to move.} We note that such a model of communication is closely related to the LOCAL model of communication studied in \cite{dereniowski2015fast}.

\bibliography{sample}

\end{document}